\documentclass[pdflatex,sn-nature]{sn-jnl}% Style for submissions to Nature Portfolio journals
\usepackage{graphicx}%
\usepackage{multirow}%
\usepackage{amsmath,amssymb,amsfonts}%
\usepackage{amsthm}%
\usepackage{mathrsfs}%
\usepackage[title]{appendix}%
\usepackage{xcolor}%
\usepackage{textcomp}%
\usepackage{manyfoot}%
\usepackage{booktabs}%
\usepackage{algorithm}%
\usepackage{algorithmicx}%
\usepackage{algpseudocode}%
\usepackage{listings}%
\usepackage{bm}
\usepackage{cleveref} 
\Crefname{figure}{Fig.}{Figures}
\usepackage{bbding}
\usepackage{makecell}
\usepackage{hyperref}
\hypersetup{
colorlinks=true,
urlcolor=blue,
}
\theoremstyle{thmstyleone}%
\theoremstyle{thmstyletwo}%

\theoremstyle{thmstylethree}%

\begin{document}

\title[Article Title]{Dark Signals in the Brain: Augment Brain Network Dynamics to the Complex-valued Field}

%%=============================================================%%
%% GivenName	-> \fnm{Joergen W.}
%% Particle	-> \spfx{van der} -> surname prefix
%% FamilyName	-> \sur{Ploeg}
%% Suffix	-> \sfx{IV}
%% \author*[1,2]{\fnm{Joergen W.} \spfx{van der} \sur{Ploeg} 
%%  \sfx{IV}}\email{iauthor@gmail.com}
%%=============================================================%%

% \author[1]{\fnm{a} \sur{Jiangnan Zhang}}\email{iauthor@gmail.com}
\author[1]{\sur{Jiangnan Zhang}}
\equalcont{These authors contributed equally to this work.}
\author[1]{\sur{Chengyuan Qian}}
\equalcont{These authors contributed equally to this work.}
\author[2,3]{\sur{Wenlian Lu}}
\author[4,5]{\sur{Gustavo Deco}}
\author*[1,2,6]{\sur{Weiyang Ding}}\email{dingwy@fudan.edu.cn}
\author*[1,2,6,7]{\sur{Jianfeng Feng}}\email{jianfeng64@gmail.com}

\affil[1]{\orgdiv{Institute of Science and Technology for Brain-Inspired Intelligence}, \orgname{Fudan University}, \orgaddress{\city{Shanghai}, \country{China}}}

\affil[2]{\orgdiv{Key Laboratory of Computational Neuroscience and Brain-Inspired Intelligence
(Fudan University)}, \orgname{Ministry of Education}, \orgaddress{\country{China}}}

\affil[3]{\orgdiv{School of Mathematical Sciences}, \orgname{Fudan University}, \orgaddress{\city{Shanghai}, \country{China}}}

\affil[4]{\orgdiv{Center for Brain and Cognition, Computational Neuroscience Group, Department of Information and Communication Technologies}, \orgname{Universitat Pompeu Fabra}, \orgaddress{\city{Barcelona}, \country{Spain}}}

\affil[5]{\orgdiv{Instituci\'{o} Catalana de Recerca i Estudis Avan\c{c}ats (ICREA)}, \orgaddress{\city{Barcelona}, \country{Spain}}}

\affil[6]{\orgdiv{MOE Frontiers Center for Brain Science}, \orgname{Fudan University}, \orgaddress{\city{Shanghai}, \country{China}}}

\affil[7]{\orgdiv{Department of Computer Science}, \orgname{University of Warwick}, \orgaddress{\city{Coventry}, \country{UK}}}

%%==================================%%
%% Sample for unstructured abstract %%
%%==================================%%

%\abstract{Brain dynamics may evolve in unobservable high-dimensional spaces, yet their modeling remains elusive. Conventional approaches typically model observed signals directly using first-order differential equations. However, Hamiltonian mechanics dictates that a complete description of a classical dynamical system requires both generalized coordinates and their conjugate momenta. Here we show that by treating fMRI signals as generalized coordinates and introducing a latent ``dark signals'' as momenta—under linearity and symmetry constraints—the Hilbert transform optimally augments the signal into the complex field, obeying a Schr\"{o}dinger-like equation. This elegant formulation provides superior data reconstruction/prediction, recapitulates hierarchical timescales, and exhibits stronger function-structure coupling compared to real-valued models. Extending to nonlinear and asymmetric regimes yields more biologically plausible effective connectivity, which notably varies with lifespan and reconfigures across cognitive states. Our framework opens new avenues in network neuroscience and provides novel insights into brain network dynamics modeling.}

\abstract{Recordings of brain activity, such as functional MRI (fMRI), provide low-dimensional, indirect observations of neural dynamics evolving in high-dimensional, unobservable spaces. Embedding observed brain dynamics into a higher-dimensional representation may help reveal functional organization, but precisely how remains unclear. Hamiltonian mechanics suggests that, by introducing an additional dimension of conjugate momenta, the dynamical behaviour of a conservative system can be formulated in a more compact and mathematically elegant manner. Here we develop a physics-informed, data-driven framework that lifts whole-brain activity to the complex-valued field. Specifically, we augment observed signals (generalized coordinates) with latent ``dark signals'' that play the role of conjugate momenta in a whole-brain Hamiltonian system. We show that the Hilbert transform provides an augmentation approach with optimal fitting accuracy within this framework, yielding a Schr\"odinger-like equation governing complex-valued, augmented brain dynamics. Empirically, this complex-valued model consistently outperforms its real-valued counterpart, improving short-horizon prediction in the linear regime (correlation 0.12$\to$0.82) and achieving superior fits under nonlinear, nonequilibrium dynamics (0.47$\to$0.88). The framework strengthens structure-function coupling, recovers hierarchical intrinsic timescales, and yields biologically plausible directed effective connectivity that varies systematically with age and reconfigures from rest to task via global rescaling plus targeted rewiring. Together, these results establish a principled, testable paradigm for network neuroscience and offer transformative insight into the spatiotemporal organization and functional roles of large-scale brain dynamics.}


%%\pacs[JEL Classification]{D8, H51}

%%\pacs[MSC Classification]{35A01, 65L10, 65L12, 65L20, 65L70}

\maketitle
\newpage
\section{Introduction}\label{sec1}
Observable activity in the natural world is often a low-dimensional projection of richer dynamics unfolding in higher dimensions. A textbook illustration is that a one-dimensional simple harmonic oscillator can be viewed as the x-projection of uniform circular motion in two dimensions \cite{openstax2016motion}. The same principle applies in neuroscience: multimodal, multi‑scale functional neuroimaging \cite{Leah2013The} provides only indirect readouts of neural activity within specific spatial units. For example, blood‑oxygen‑level‑dependent (BOLD) signals recorded by functional magnetic resonance imaging (fMRI) reflect vascular and metabolic processes coupled to neuronal activity, rendering it a delayed, low-frequency proxy rather than a direct measure \cite{ZHANG2020116390}. Fitting models purely within the observed space is therefore insufficient to uncover underlying mechanisms, underscoring the need to augment measured dynamics with latent degrees of freedom.

Emerging evidence suggests that high-dimensional representations, such as complex-valued modeling, facilitate the elucidation of collective oscillations within brain networks \cite{bib3,bib22,bib47,bib19,bib1,bib21,bib20,bib56,bib57} and identify fundamental activity patterns \cite{bib44,bib32,bib81}. One line of work uses Stuart–Landau (Hopf) oscillators to model hidden neural dynamics in the complex-valued field and to generate synthetic data matching empirical features such as functional connectivity and frequency structure \cite{2024Human,bib80}. A complementary line lifts real measurements to analytic (complex-valued) signals via the Hilbert transform and then extracts organizing principles with methods like Complex PCA (CPCA) \cite{bib44} and Complex Harmonic Decomposition (CHARM) \cite{bib32,bib81}. Despite these advances, a unified framework that both augments spatiotemporal brain dynamics into a higher-dimensional space and explains their organizational principles and functional significance within a single, physics-grounded formulation remains lacking.

Hamiltonian mechanics offers such a foundation that any conservative system can be specified by a conjugate pair of generalized coordinates and momenta that follow Hamilton's equations \cite{1983Mathematical,2015An}. Inspired by this, we treat node‑wise BOLD time series as generalized coordinates and introduce their conjugate momenta as latent variables, termed ``dark signals'', together forming a whole‑brain Hamiltonian system. We demonstrate that, within this framework, the Hilbert transform provides the optimal augmentation to construct ``dark signals'', yielding a Schr\"{o}dinger-like equation in the complex-valued field.

Building on this theoretical basis, we develop a data‑driven framework that models complex-valued analytic signals in brain networks via the Schr\"{o}dinger-like equation. This framework embeds the original dissipative system into a higher-dimensional space where the combined dynamics are conservative/Hamiltonian. Applying this framework to the 3T Human Connectome Project (HCP) dataset \cite{hcp2017manual}, we show that the Schr\"{o}dinger-type model enables more precise characterization of empirical dynamics than their real-valued counterpart ({\Cref{tab1}}). Using the complex-valued coupling matrix, we identify stronger structure-function couplings than those using functional connectivity and the real-valued model. When stimulating the primary sensory cortices, we successfully reproduce the widely observed hierarchy of intrinsic timescales, whereas the real-valued model fails to capture this hallmark feature. Furthermore, extending the framework to nonlinear, nonequilibrium regimes yields a novel effective connectivity. The resulting network organization is more biologically plausible, shows systematic variation across the lifespan in HCP-D/YA/A (ages 8–100), and reconfigures under task demands via global rescaling of resting-state couplings combined with targeted rewiring that preserves propagation backbones while amplifying task-specific receivers. Taken together, this study provides novel insights into brain dynamics modeling, establishes a testable analytic paradigm for network neuroscience research, and demonstrates its functional significance.

% \begin{table}[h]
% \caption{Performance Comparison Between Complex-Valued and Real-Valued Models}\label{tab2}
% \centering
% \begin{tabular*}{1\textwidth}{@{\extracolsep\fill}lcccc}
% \toprule%
% & \multicolumn{4}{@{}c@{}}{Linear Model}
% \\\cmidrule{2-5}
% & \makecell{Reconstruction \footnotemark[1] }
% & \makecell{Short-horizon \\ Prediction \footnotemark[2]}  
% & \makecell{Timescale  Hierarchy} 
% &  \makecell{Structure-function \\ coupling (SFC) } \\
% \midrule
% Complex  & $0.93\pm 0.03$ & $0.82\pm 0.02$ & \Checkmark & $ 0.35 \pm 0.14$ \\
% Real   & $0.89\pm 0.04$ & $0.12\pm 0.34$  & \XSolidBrush & $0.10 \pm 0.08$ \\
% \midrule
% & \multicolumn{4}{@{}c@{}}{Nonlinear Model}
% \\\cmidrule{2-5}
% & \makecell{Fitting  Accuracy} 
% & \makecell{Biological \\ Plausibility } 
% & \makecell{Communicability }
% & \makecell{Information \\ Transmission}\\
% \midrule
% Complex  & $0.88\pm 0.04$  & Better & $ 7.10\times 10^{-3}\pm 3.57 \times 10^{-5}$ & Effective\\
% Real   & $0.47\pm 0.12$ &  Worse & $ 6.70\times 10^{-3}\pm 1.78 \times 10^{-4}$ & Ineffective\\
% \botrule
% \end{tabular*}
% \footnotetext[1]{Reconstruction accuracy of 100 TRs time series.}
% \footnotetext[2]{Prediction accuracy of 10 TRs time series.}
% \end{table}

\begin{table}[h]
\caption{Performance comparison between the Schr\"{o}dinger-type complex-valued model and its real-valued counterpart. Comprehensive linear model comparisons are provided in Sec. \ref{2.4}, while detailed nonlinear model analyses are presented in Sec. \ref{schrodinger}. }\label{tab1}
\begin{tabular*}{\textwidth}{@{\extracolsep\fill}lcccccc}
\toprule%
& \multicolumn{3}{@{}c@{}}{Linear Model} & \multicolumn{3}{@{}c@{}}{Nonlinear Model} \\\cmidrule{2-4}\cmidrule{5-7}%
& \makecell{Short-\\horizon \\ Prediction\dag}  
& \makecell{Timescale \\ Hierarchy} 
&  \makecell{Structure- \\ Function \\ Coupling } 
& \makecell{Fitting \\ Accuracy} 
& \makecell{Biological \\ Plausibility } 
& \makecell{Effective \\ Information \\ Transfer} \\
\midrule
Complex  & $0.82\pm 0.02$* & \Checkmark & $ 0.35 \pm 0.14$*& $0.88\pm 0.04$* & \Checkmark & \Checkmark\\
Real   & $0.12\pm 0.34$* & \XSolidBrush & $0.10 \pm 0.08$*& $0.47\pm 0.12$* &  \XSolidBrush  & \XSolidBrush\\
\botrule
\end{tabular*}
\footnotetext[\dag]{Prediction accuracy of 10 TRs resting-state fMRI data.}
\footnotetext[*]{Pearson's correlation, mean ± s.d.}
\end{table}

\section{Results}\label{sec2}

\subsection{Theoretical framework of brain dynamics modeling in higher-dimensional space}
\begin{figure*}[htbp]
	\centering
	\includegraphics[width=1\textwidth]{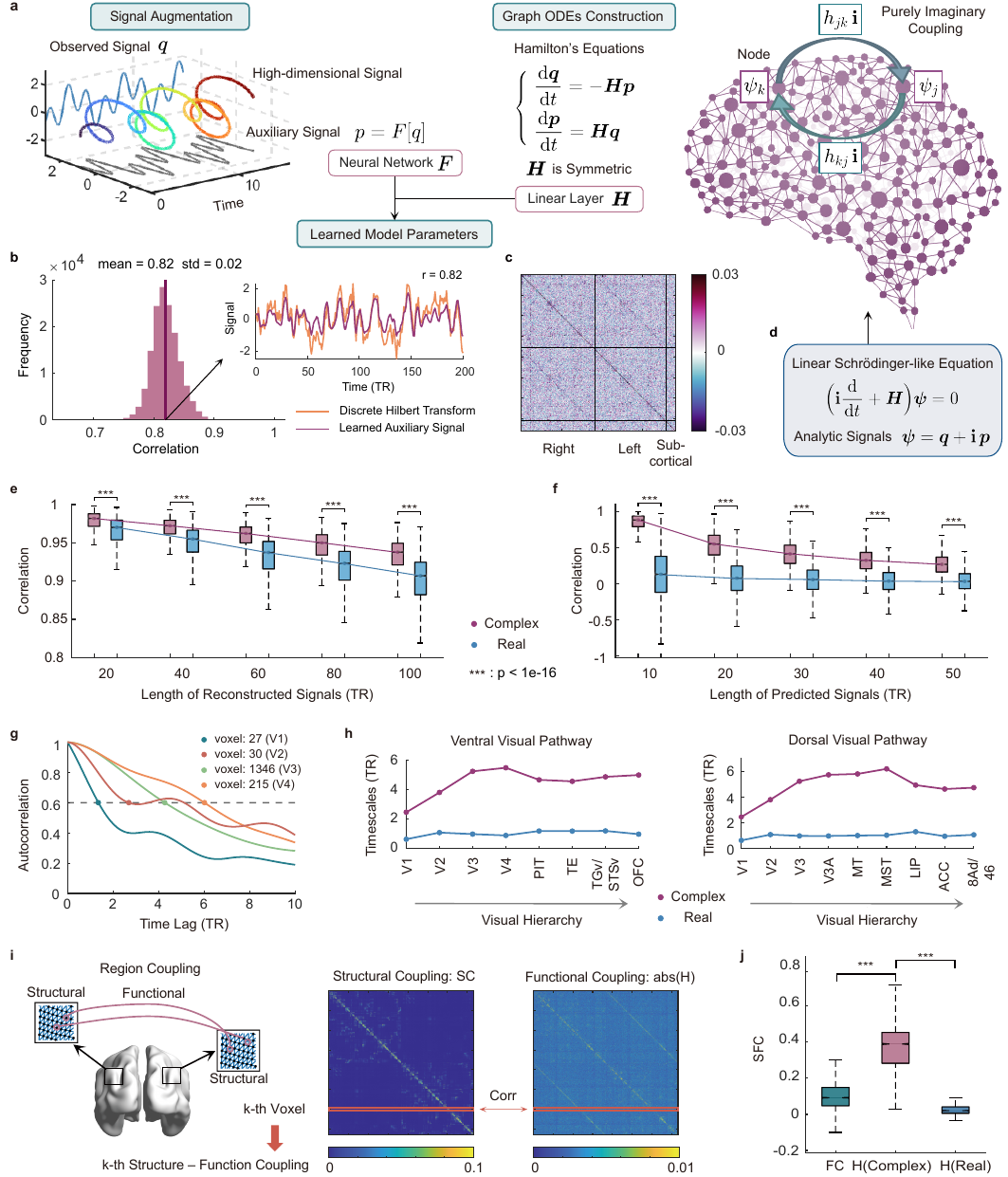} %
    	\caption{Modeling brain dynamics using a linear Schr\"{o}dinger-like equation in the complex-valued field. $\bm{\mathrm{a}}$, Brain dynamics are modeled using Hamilton's equations, with a coupling parameter $\bm{H}$ and auxiliary signals estimated through neural networks. This framework is applied to fMRI data of 395 participants. $\bm{\mathrm{b}}$, The learned auxiliary signals exhibit high correlation with the discrete Hilbert transform of observed signals. $\bm{\mathrm{c}}$, Learned $\bm{H}$ is a positive definite matrix. $\bm{\mathrm{d}}$, These findings support modeling brain dynamics using analytic signals within the linear Schr\"{o}dinger-like framework. $\bm{\mathrm{e-f}}$, The linear Schr\"{o}dinger-like framework outperforms its real-valued counterpart in reconstructing and predicting neural activity. $\bm{\mathrm{g}}$, Autocorrelation of response signals following V1 stimulation. $\bm{\mathrm{h}}$, Visual cortical exhibits hierarchical intrinsic timescales along ventral and dorsal pathways. The above patterns are undetectable by real-valued models. $\bm{\mathrm{i}}$, SFC quantification schematics. $\bm{\mathrm{j}}$, Schr\"{o}dinger-derived $\bm{H}$ provides superior structure-function mapping ability compared to the real-valued connectivity.} 
	\label{fig:Linear}
\end{figure*}

To model brain dynamics in higher dimensions, we propose a whole-brain Hamiltonian framework that augments the observed brain signal with an auxiliary signal (latent ``dark signals'')(\Cref{fig:Linear}$\bm{\mathrm{a}}$). Inspired by Hamiltonian mechanics \cite{1983Mathematical,2015An}, we regard the brain as an $N$-dimensional Hamiltonian system, embed the interregional coupling matrix $\bm{H}$ into the quadratic Hamiltonian, and describe each node by a ``coordinate-momenta'' pair $(q_k,p_k)$. The observed brain signals $\bm{q}(t) = [q_1,\dots,q_N]^\top$ play the role of generalized coordinates, while the latent auxiliary signals $\bm{p}(t) = [p_1,\dots,p_N]^\top$ serve as the corresponding conjugate momenta and are modeled via a transformation of the observations, $\bm{p}(t) = F[\bm{q}](t)$.

\begin{equation}\label{2.1-1}
\begin{aligned}
 \frac{{\rm d} \bm{q}}{{\rm d} t} =  -\bm{H} \bm{p}, \;
 \frac{{\rm d} \bm{p}}{{\rm d} t}  =\bm{H} \bm{q}.\\ 
\end{aligned}
\end{equation}
This construction follows the standard Hamiltonian formulation in which dynamics are expressed in the $2N$-dimensional phase space $(\bm{q},\bm{p})$. Assuming that the coupling matrix is symmetric, we develop a neural network (Sec. \ref{Hilbert}) to learn auxiliary signals and the interregional coupling from functional magnetic resonance imaging (fMRI) data released by the Human Connectome Project (HCP) \cite{bib42}. Notably, the learned auxiliary signals have a high correlation with the discrete Hilbert transform \cite{Horel1984ComplexPCA,2014Survey} of observed signals ($r=0.82\pm 0.02$, mean $\pm$ s.d.; Pearson’s correlation; \Cref{fig:Linear}$\bm{\mathrm{b}}$). And the estimated coupling matrix $\bm{H}$ is a positive-definite matrix with a three-diagonal structure, showing a biologically plausible connectivity among functional regions (\Cref{fig:Linear}$\bm{\mathrm{c}}$). Combined with theoretical proof (Supplementary Information S1), we indicate that, under symmetric and positive-definite constraints of the coupling matrix, the Hilbert transform is the optimal choice for signal augmentation that satisfies the whole-brain Hamiltonian system \cite{2023Different}.

Based on the above experimental and theoretical results, we directly view the Hilbert transform of the observed signals as ``dark signals'' to model the brain's dynamical system. Hence, the Hamiltonian framework \eqref{2.1-1} can be reformulated into a complex-valued representation: the complex-valued analytic signals $\bm{\psi} = \bm{q} + {\bf i} \bm{p} $ obey a linear Schr\"{o}dinger-type evolution on the brain network,
\begin{equation}\label{linear_schrodinger}
    \Big({\bf i} \frac{{\rm d}}{{\rm d} t} + \bm{H}\Big) \bm{\psi} = 0.
\end{equation}

This complex-valued model admits an explicit-form solution $\bm{\psi}(t) = \exp(\bm{H}t)\bm{\psi}(0)$, whose energy function $E(\bm{\psi}) = \bm{\psi}^* \bm{H} \bm{\psi}$ is a constant during evolution.
That is, by modeling brain dynamics in this form, we regard observed brain signals as a partial measurement of a higher-dimensional conservative system, which differs from existing paradigms for brain dynamics modeling \cite{bib16,bib51,bib57}.

Although Hamilton's equations typically describe conservative systems, this does not imply that the brain's dynamical system is in an equilibrium state. In fact, the inter-regional coupling in the linear Schr\"{o}dinger-like equation \eqref{linear_schrodinger} exhibits antisymmetry, enabling our complex-valued linear model to capture the irreversibility of brain dynamics \cite{2024Mind,deco2022insideout} (Supplementary Information S4). This model also reveal a novel coupling mechanism: the coupling between nodes in the complex-valued field is achieved via a purely imaginary effect (\Cref{fig:Linear}$\bm{\mathrm{d}}$), elucidating the physical significance of both observed and auxiliary signals.

\subsection{Characterize large-scale brain organization using a linear Schr\"{o}dinger-like equation} \label{2.4}
We first validate the effectiveness of the Schr\"{o}dinger-like framework in characterizing brain organization compared with conventional real-valued models. We train this model using a segment of voxel-level resting-state analytic signals spanning 300 TRs, and subsequently provide the trained model with the signals at first and the $301^{st}$ TR as initial states to perform reconstruction and prediction, respectively (Supplementary Information S3). As a control, we train a real-valued model using the same procedure. The complex-valued model achieves higher reconstruction accuracy, with Pearson's correlation reaching up to 0.85 for all voxels within 100 TRs (\Cref{fig:Linear}$\bm{\mathrm{e}}$). Importantly, it also demonstrates superior generalization capability compared to its real-valued counterpart (Complex: $r=0.84\pm0.14 $, mean $\pm$ s.d.; Real: $r=0.12\pm 0.34$, mean $\pm$ s.d.; Pearson’s correlation at 10 TR; \Cref{fig:Linear}$\bm{\mathrm{f}}$). Moreover, we provide the trained model with a synthetic initial state to simulate a stimulus to the V1 area in the left hemisphere. The generated stimulus-evoked dynamics exhibit propagation characteristics, with the earlier visual cortex (e.g. V2, V3) exhibits a stronger and earlier response and rapid decay in contrast to the higher-order cortical regions  (Supplementary Video S13). Furthermore, we quantified intrinsic timescales by calculating the autocorrelation function of response signals (\Cref{fig:Linear}$\bm{\mathrm{g}}$, Sec.\ref{timescale}), revealing a timescale hierarchy \cite{bib66,bib67,bib68} across two visual pathways (\Cref{fig:Linear}$\bm{\mathrm{h}}$). In contrast, the stimulus-evoked dynamics in the real-valued linear model fail to reveal this hierarchical organization. These findings reveal that upon introducing the auxiliary signal, the linear model is adequate to depict large-scale spontaneous brain dynamics.

\begin{figure*}[htbp]
	\centering
	\includegraphics[width=1\textwidth]{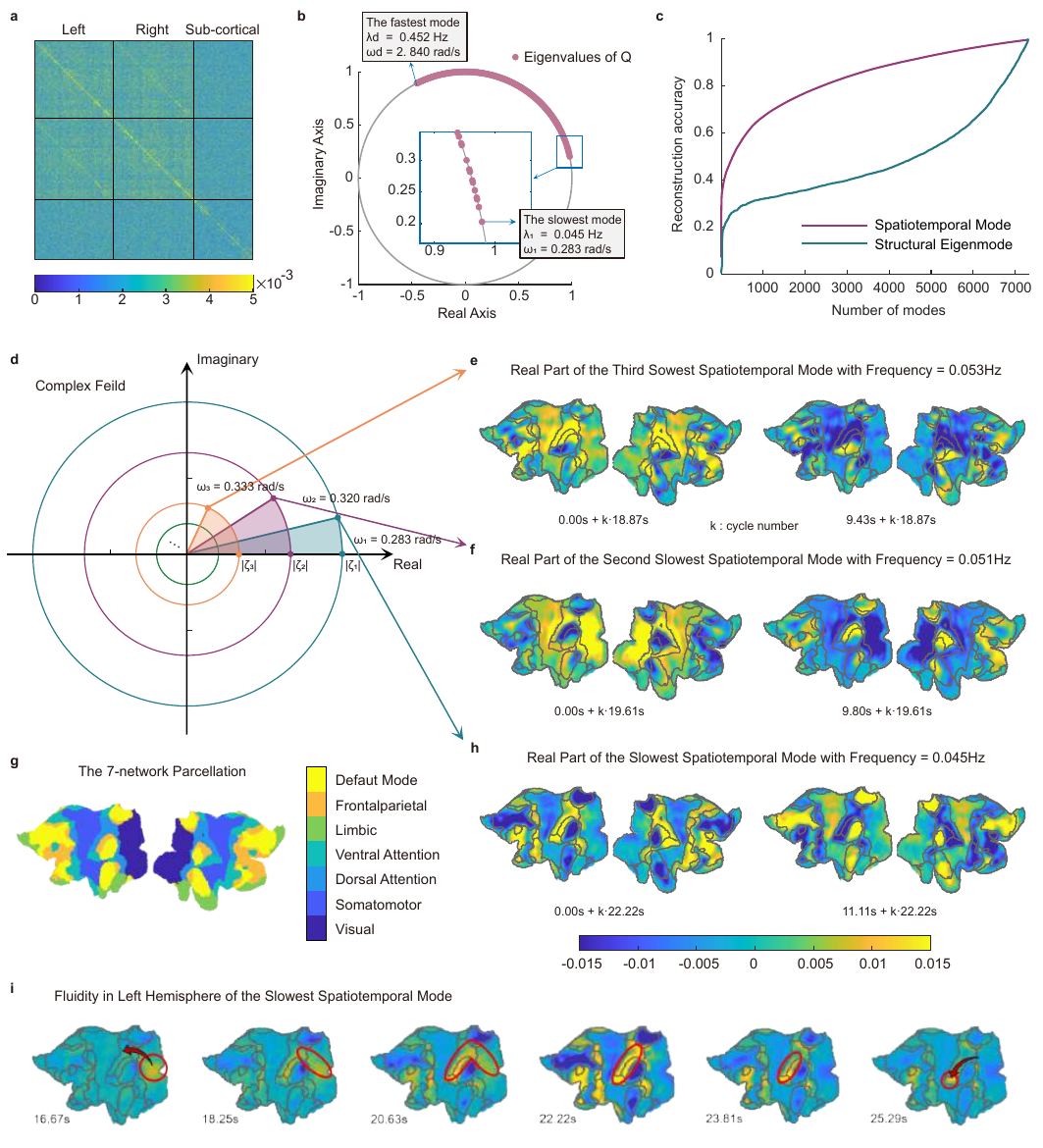} %
 	\caption{Form and properties of three prominent Schr\"{o}dinger-derived spatiotemporal modes. $\bm{\mathrm{a}}$, The group-level transfer matrix $\bm{Q}$ is tri-diagonal with large diagonal elements. $\bm{\mathrm{b}}$, The eigenvalues of $\bm{Q}$ indicate mode frequencies ranging from 0.045 to 0.452 Hz. $\bm{\mathrm{c}}$, Schr\"{o}dinger-derived spatiotemporal modes outperform structural eigenmodes in reconstructing rs-fMRI data. $\bm{\mathrm{d}}$, Energy-state stratification depends on the amplitude $\zeta$: low amplitudes correspond to the low-energy orbits, while high amplitudes reflect elevated energy states. $\bm{\mathrm{e}}$-$\bm{\mathrm{h}}$, Real-component dynamics of three fundamental low-frequency modes. $\bm{\mathrm{g}}$, This brain map shows the parcellation of 7 functional networks on the cerebral cortex. $\bm{\mathrm{h}}$, The slowest mode accurately identifies these functional networks. Grey lines delineate the 7 functional network boundaries. $\bm{\mathrm{i}}$, Cortical wave dynamics are visualized by the slowest mode, propagating from visual network toward the ventral and dorsal attention networks.} 
	\label{fig:mode}
\end{figure*}

We next investigate the global functional brain organization within the framework of the linear Schr\"{o}dinger equation. Dynamic mode decomposition \cite{bib27} is applied to the group-level transfer matrix (The exponent of the coupling matrix, \Cref{fig:mode}$\bm{\mathrm{a}}$, Sec. \ref{spatio}), identifying complex harmonics at various frequencies (ranges from 0.045 Hz to 0.452 Hz; \Cref{fig:mode}$\bm{\mathrm{b}}$). These frequencies correspond to distinct energy orbits (\Cref{fig:mode}$\bm{\mathrm{d}}$). Each spatiotemporal mode describes brain dynamics on a specific energy orbit. Spontaneous brain dynamics in the complex-valued field can be regarded as a weighted sum of different modes (Sec.\ref{spatio}). Guided by this decomposition, we evaluate the accuracy of these spatiotemporal modes in capturing resting-state brain activity. Our results demonstrate that spatiotemporal modes significantly outperform structural eigenmodes \cite{bib34,bib35,bib45,bib46}, which are derived from structural connectivity matrices (Sec. \ref{structural_eigenmodes}), in reconstructing rs-fMRI data (t-value = $128.83$, $p < 10^{-16}$, paired t-test; \Cref{fig:mode}$\bm{\mathrm{c}}$; \ref{sec_FC_recon}). The reconstruction accuracy is quantified by the correlation between empirical and spontaneous FC matrices.

We extract the three slowest-varying spatiotemporal modes (Supplementary Video S12) to characterize the principal components of resting-state dynamics in the complex-valued field. The primary spatiotemporal mode has the slowest variation, with an oscillatory frequency of 0.045 Hz (\Cref{fig:mode}$\bm{\mathrm{h}}$). Its real part's dynamics exhibits a remarkable alignment with the 7-network gradient-based parcellation \cite{bib28} (\Cref{fig:mode}$\bm{\mathrm{g}}$). The initial dynamic phase is characterized by strong negative BOLD amplitudes in the default mode network (DMN), contrasting with concurrent positive BOLD amplitudes in the frontoparietal, dorsal attention, and ventral attention networks. This pattern undergoes a complete amplitude reversal after 11.11 seconds, with DMN exhibiting positive amplitudes while the frontoparietal and attention networks showed negative amplitudes. After an equivalent duration (11.11 s), the dynamics reverted to its initial stage, completing a full oscillatory cycle. This mode also reveals a wave fluidity propagating from the visual network to the dorsal and ventral attention networks (\Cref{fig:mode}$\bm{\mathrm{i}}$). The secondary spatiotemporal mode, which is the second slowest-varying mode, exhibited an oscillatory frequency of 0.051 Hz (\Cref{fig:mode}$\bm{\mathrm{f}}$). This mode demonstrated a significant anti-correlated oscillation between the lower-order functional cortex (visual, auditory) and both the frontoparietal as well as the parietal cortex. The tertiary (third slowest-varying) spatiotemporal mode oscillated at 0.053 Hz, also displaying robust anti-correlated dynamics across the human cortex (\Cref{fig:mode}$\bm{\mathrm{e}}$).

We further reveal that the Schr\"{o}dinger model provides a valid underlying structure for brain dynamics and functions. Hence, we seek to explore structure-function coupling (SFC) using the linear Schr\"{o}dinger equation. Understanding how the brain's anatomical structure gives rise to a wide range of complex functions remains a fundamental and unresolved challenge in neuroscience \cite{bib38}. The literature has primarily focused on examining the macroscale coupling between structural (SC) and functional connectivity (FC). We quantify SFC by the correlation coefficient between corresponding rows of the SC and the FC (\Cref{fig:Linear}$\bm{\mathrm{i}}$) \cite{bib37,bib39,bib40}. Notably, the SFC calculated using our complex-valued coupling matrix exhibits significantly stronger values in most regions compared to the real-valued connectivity (FC: SFC $ = 0.102 \pm 0.075$; estimated real-valued coupling matrix: SFC $ = 0.028 \pm 0.032$), with a range from -0.020 to 0.721 (\Cref{fig:Linear}$\bm{\mathrm{j}}$). This finding suggests that coupling in the complex-valued field captures more structural properties of the cerebral cortex than that in the real field. In summary, these findings demonstrate that the linear Schr\"{o}dinger equation framework offers a more effective approach for characterizing large-scale spatiotemporal brain dynamics, providing novel insights into the modeling of brain activity and the investigation of brain function and structure. 

\subsection{A nonlinear data-driven paradigm also reveals Schr\"{o}dinger-like dynamics in the brain}\label{schrodinger}

\begin{figure*}[htbp]
	\centering
	\includegraphics[width=1\textwidth]{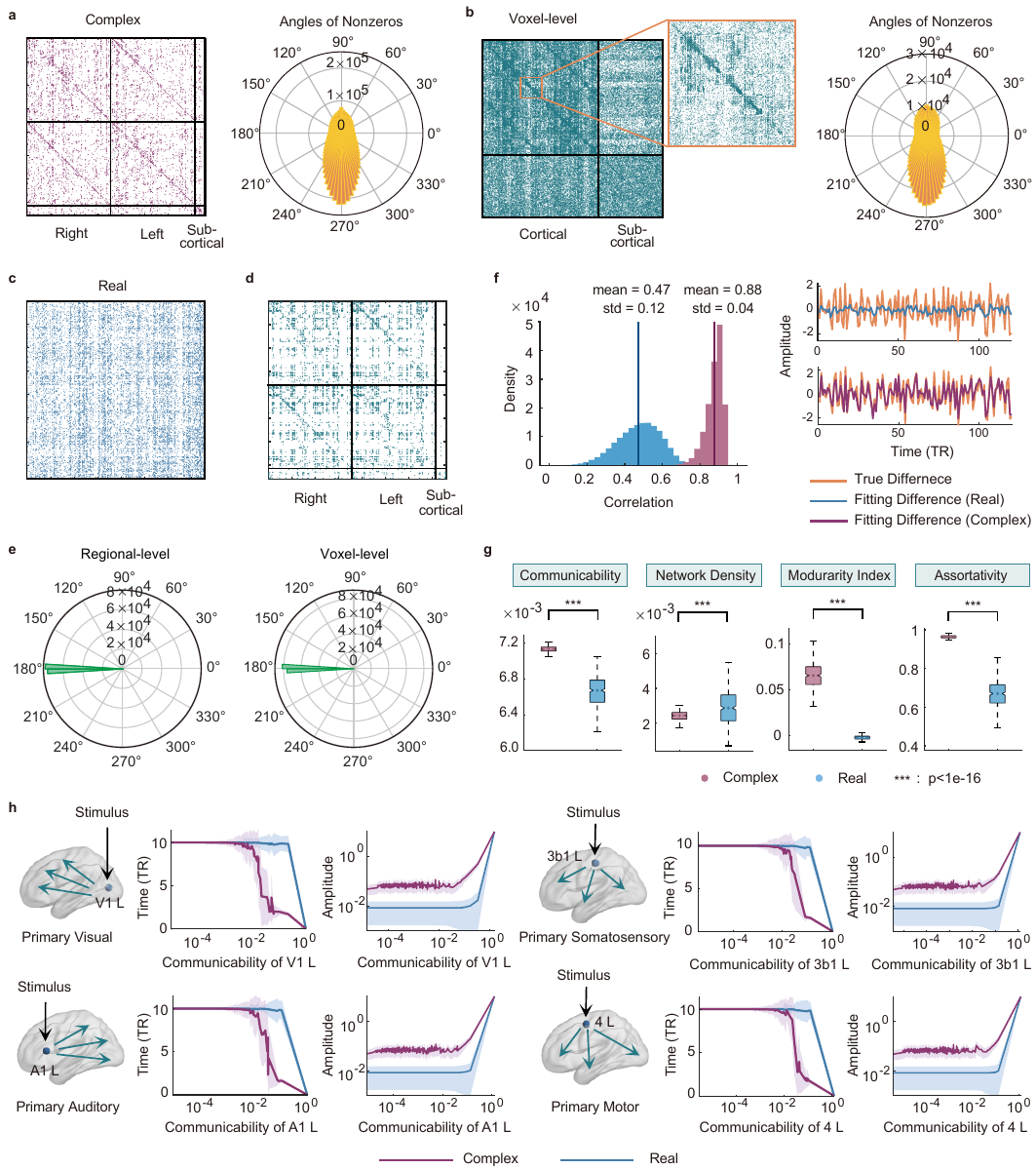} %
	\caption{Data-driven nonlinear modeling of brain dynamics in the complex-valued field. $\bm{\mathrm{a-c}}$, The complex-valued nonlinear model is applied to analytic signals at different spatial scales: $\bm{\mathrm{a}}$ regions and $\bm{\mathrm{b}}$ voxels, revealing a nonlinear Schr\"{o}dinger-like governing equation. While the real-valued coupling matrix shows no significant structure, the complex-valued coupling matrix is asymmetric and identifies strong functional connectivity between homologous regions across hemispheres. $\bm{\mathrm{d}}$, Granger causality analysis of rs-fMRI identifies a strong bidirectional interhemispheric causal relationship. $\bm{\mathrm{e}}$, The calculated Hamiltonian of each time is non-Hermitian. $\bm{\mathrm{f}}$, The complex-valued model exhibits higher fitting accuracy than its real-valued counterpart. $\bm{\mathrm{g}}$, Network measures indicate a significant statistical divergence between complex-valued and real-valued networks. The complex-valued model has superior communicability, sparser network density, more precise functional segregation, and stronger assortativity than the real-valued one. $\bm{\mathrm{h}}$, Under structural connectivity constraints, complex-valued brain dynamics enable efficient signal propagation across anatomically communicable regions, which cannot be captured by real-valued dynamics. (Note: Amplitude and communicability are presented on logarithmic scales.)} 
	\label{fig:Nonlinear}
\end{figure*}

Beyond linear constraints, we extend our framework to a nonlinear regime for modeling whole-brain complex-valued analytic signals. Our complex-valued parametric model is designed as a generic form, with nonlinear polynomial evolution and linear interregional coupling (Sec.\ref{data-driven}). Assuming that the brain connectivity exhibits sparsity \cite{bib69}, model parameters are estimated using a sparse optimization approach (Extended Data \Cref{fig:7}$\bm{\mathrm{a}}$). We apply it to the resting-state analytic signal, which is derived by augmenting the observed BOLD signal to the complex-valued field through the Hilbert transform. Notably, the estimated complex-valued coupling elements are highly imaginary at various spatial scales, with their angles exhibiting a significant bimodal distribution characterized by two prominent peaks at $\pi/2$ ($90^{\circ}$) and $-\pi/2$ ($270^{\circ}$) (regional level: $R=0.38$, $p=0$, \Cref{fig:Nonlinear}$\bm{\mathrm{a}}$; voxel level: $R=0.36$, $p=0$, \Cref{fig:Nonlinear}$\bm{\mathrm{b}}$; Rayleigh test after angle doubling, Supplementary Information S9). We validate this finding using fMRI data from the UK biobank dataset \cite{2015UK} and obtain the same result (Extended Data \Cref{fig:surrogate}). However, the coupling matrices derived from the surrogate data and the NYC traffic data \cite{nyc_taxi_2015} are not purely imaginary (Supplementary Information S9). This result elucidates a novel mechanism: Functional coupling between brain regions is accomplished by the mutual interaction between auxiliary signals and observed signals. By approximating coupling elements as purely imaginary numbers, the governing equation of complex-valued brain dynamics suggests structures similar to a nonlinear Schr\"{o}dinger-like equation on the graph \eqref{nonlinear_schrodinger}, where  $\bm{H}  \in \mathbb{R}^{N \times N}$ and $q(\bm{\psi})$ is the nonlinear self-coupling term.

\begin{equation}\label{nonlinear_schrodinger}
    \Big({\bf i} \frac{{\rm d}}{{\rm d} t} + \bm{H}\Big) \bm{\psi} = q(\bm{\psi}).
\end{equation} 

Furthermore, the observed coupling asymmetry fundamentally indicates the system's deviation from thermodynamic equilibrium, which aligns well with the intrinsic non-equilibrium characteristics of brain dynamics \cite{2013Nonequilibrium,2024Mind}. The coupling asymmetry characterizes the directional information flow within the whole-brain network, with asymmetric interactions being particularly prominent in subcortical-cortical connections (\Cref{fig:Nonlinear}$\bm{\mathrm{a-b}}$, Extended Data \Cref{fig:surrogate}). This directional coupling pattern exhibits high consistency with Granger causal networks ($r = 0.33$, $p=0$, Pearson’s correlation; \Cref{fig:Nonlinear}$\bm{\mathrm{d}}$). Such coupling directionality leads to the non-unitary properties of temporal evolution operators, revealing the time irreversibility embedded in neural dynamics \cite{2022Temporal,2024Broken}. The non-Hermitian Hamiltonian (\Cref{fig:Nonlinear}$\bm{\mathrm{e}}$) governing this process intrinsically reflects the intrinsic dissipative properties of neural systems and provides a robust theoretical foundation for understanding the metastable criticality observed in brain dynamics \cite{bib51,bib79}. Therefore, compared to conservative and equilibrium Hamiltonian systems, the nonlinear Schr\"{o}dinger-like model can more effectively capture the non-equilibrium properties of brain dynamics, providing a more precise theoretical foundation for understanding brain dynamical behaviors.

Considering a more biologically structure-informed coupling, we constrain the coupling using the structural connectivity (SC) matrix, with a complex-valued global coupling parameter $g$ (Sec. \ref{SC_constraint}). Employing the same data-driven paradigm to fit analytic signals, we derive the consistent conclusion that the estimated global coupling parameter $g$ for all individuals is highly imaginary, with their angles exhibiting a significant unimodal distribution, peaking at $-\pi/2$ ($R=1.00$, $p=0$, Rayleigh test, Supplementary Fig. 5$\bm{\mathrm{a}}$). This result further supports the utility of Schr\"{o}dinger-type equations for modeling large-scale brain activity in the complex-valued field.

\subsection{Complex-valued nonlinear models benchmark against real-valued ones}\label{complex-real comparison}

To systematically evaluate the complex-valued nonlinear modeling, we compare the solved models and corresponding dynamics under the real and complex representations. First, we utilize the proposed generic data-driven paradigm to construct both real-valued and complex-valued models. The complex-valued model exhibits outstanding performance in fitting the data ($r=0.88\pm0.04$, mean $\pm$ standard deviation (s.d.), Pearson’s correlation; \Cref{fig:Nonlinear}$\bm{\mathrm{f}}$), outperforming the real-valued model ($r=0.47\pm0.12$, mean $\pm$ s.d.). Comparing their coupling structure, the estimated real-valued coupling matrix lacks a modular structure and fails to represent the strong coupling between corresponding functional regions in contralateral hemispheres (\Cref{fig:Nonlinear}$\bm{\mathrm{a}},\bm{\mathrm{c}}$). 

To quantify the capability of estimated network structures, we employ several network measures from graph theory \cite{bib77,bib78}. The communicability of complex-valued networks is significantly higher than real-valued ones ($t$-value $=50.74$, $p=2.63\times10^{-249}$, paired t-test; \Cref{fig:Nonlinear}$\bm{\mathrm{g}}$), suggesting that information transmission within complex-valued networks is more efficient. Additionally, under the same sparsity constraint, the network density of complex-valued ones is lower ($t$-value $ =-9.92$, $p=6.12\times10^{-22}$, paired t-test), and the modularity index is higher ($t$-value $ =94.65$, $p=0$, paired t-test). Complex-valued networks exhibit higher assortativity, providing insights into the preferential attachment of brain regions ($t$-value $ =80.71$, $p=0$, paired t-test). These results indicate that the complex-valued network structure balances specialization and integration, optimizing the brain's ability to process information efficiently.

To further demonstrate that this superiority is not simply due to a better-fitted network structure, we employ the SC as the interregional coupling in both models and conduct stimulation experiments (Sec.\ref{SC_constraint}). We stimulate the primary visual, auditory, somatosensory, and motor cortex, calculating the timing and amplitude of the first peak of response signals in other brain regions within 10 TRs (Sec.\ref{stimulus-evoked}). The complex-valued brain dynamics demonstrate a remarkable ability for whole-brain transmission, enabling signals to pass through regions with structural barriers (\Cref{fig:Nonlinear}$\bm{\mathrm{h}}$, Supplementary Video S13). In anatomically communicable regions, responses evoked by stimulation are stronger ($t$-value$=-63.40\sim-29.67$, $p_{FDR}< 8.87\times 10^{-126}$, paired t-test) and occur earlier ($t$-value$=7.92\sim169.45$, $p_{FDR}< 9.30\times 10^{-15}$, paired t-test) in complex-valued models compared to real-valued ones. For example, when the primary visual region (V1) is stimulated in the real field, only the visual cortex activates, with response amplitudes lower than $0.036$. In contrast, in the complex-valued field, signals propagate along the ventral and dorsal pathways of hierarchical visual processing \cite{bib64,bib65} to high-order regions within 10 TRs, exhibiting significantly higher response amplitudes ranging from $0.045$ to $0.472$. In summary, these results reveal that introducing latent ``dark signals'' to brain dynamics allows capturing functional characteristics and effective information flow imperceptible to real-valued models.

\subsection{Complex-valued connectivity adapts throughout the human lifespan}\label{age}

\begin{figure*}[htbp]
	\centering
	\includegraphics[width=1\textwidth]{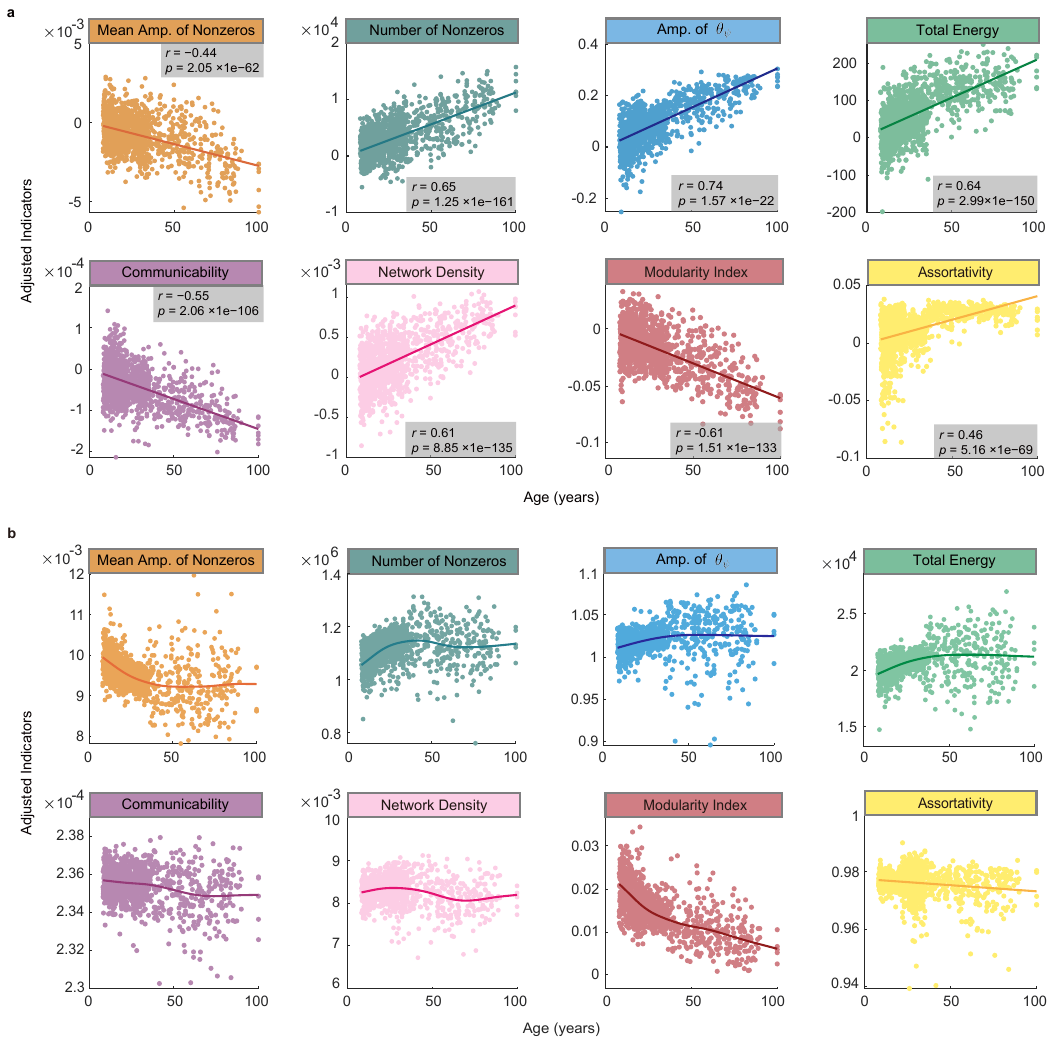} %
	\caption{Age-related changes in complex-valued functional coupling at different spatial scales.  $\bm{\mathrm{a}}$, Regional-level analysis after correcting for age-group biases (linear regression) shows significant age correlations in all coupling measures. $\bm{\mathrm{b}}$, At the voxel level, a generalized additive model is fitted separately for each indicator to capture age-related changes in whole-brain coupling. Whole-brain coupling patterns before adulthood align with the regional level but stabilize in later life.} 
	\label{fig:Age}
\end{figure*}

This data-driven paradigm provides an effective and efficient way to characterize the directed effective connectivity of the brain, facilitating further investigation in large-scale datasets. Therefore, we next explore the human lifespan changes in complex-valued functional coupling. We estimate coupling matrices using resting-state fMRI data from HCP-Development \cite{bib72} ($n=633$), HCP-Young Adults \cite{bib43} ($n=395$), and HCP-Aging \cite{bib71} ($n=293$) projects, spanning a wide age range ($8-100$ years old). To summarize the essential characteristics of the coupling, we compute the number, mean amplitude of nonzero elements (without the diagonal), and the aforementioned network measures. 

At the regional level, the lifespan changes of these features exhibit a significant linear tendency (\Cref{fig:Age}$\bm{\mathrm{a}}$). The number of nonzero elements in the coupling matrix increases throughout the lifespan ($r = 0.65, p<0.001$, Pearson’s correlation), while the average coupling strength shows a statistically significant decrease  ($r = -0.44, p<0.001$). These results are consistent with previous studies showing that the human brain initially exhibits relatively sparse interregional connectivity during development, followed by a period of densification in which the number of connections gradually increases \cite{bib69,bib70}. The self-coupling strength, quantified by the coefficient amplitude of $\bm{\psi}$, exhibits a significant positive correlation with age ($r = 0.74, p<0.001$). Similarly, the absolute value of the Hamiltonian, which denotes the total energy of the system \cite{bib26}, demonstrates substantial age-related increases ($r = 0.64, p<0.001$). Throughout the lifespan, both communicability and modularity indices show significant declining trends ($r = -0.55, p<0.001$ and $r = -0.61, p<0.001$, respectively). The network density and assortativity exhibit robust positive associations with age ($r = 0.61, p<0.001$ and $r = 0.46, p<0.001$, respectively). At the voxel level, these changes exhibit a nonlinear tendency. Hence, we calculate the same coupling indicators and model their age-dependent variations using generalized additive models (GAMs) (\Cref{fig:Age}$\bm{\mathrm{b}}$, Supplementary Information S8). These developmental changes observed during adolescence and middle age align with variations at the regional level. However, divergence occurs in advanced age, potentially driven by senescence-related alterations in fine-scale intraregional coupling. These findings yield novel insights into age-related changes in brain connectivity and demonstrate the functional relevance of the complex-valued functional coupling.

\subsection{Task-evoked changes in the complex-valued connectivity exhibit a low-rank structure}

\begin{figure*}[htbp]
	\centering
	\includegraphics[width=1\textwidth]{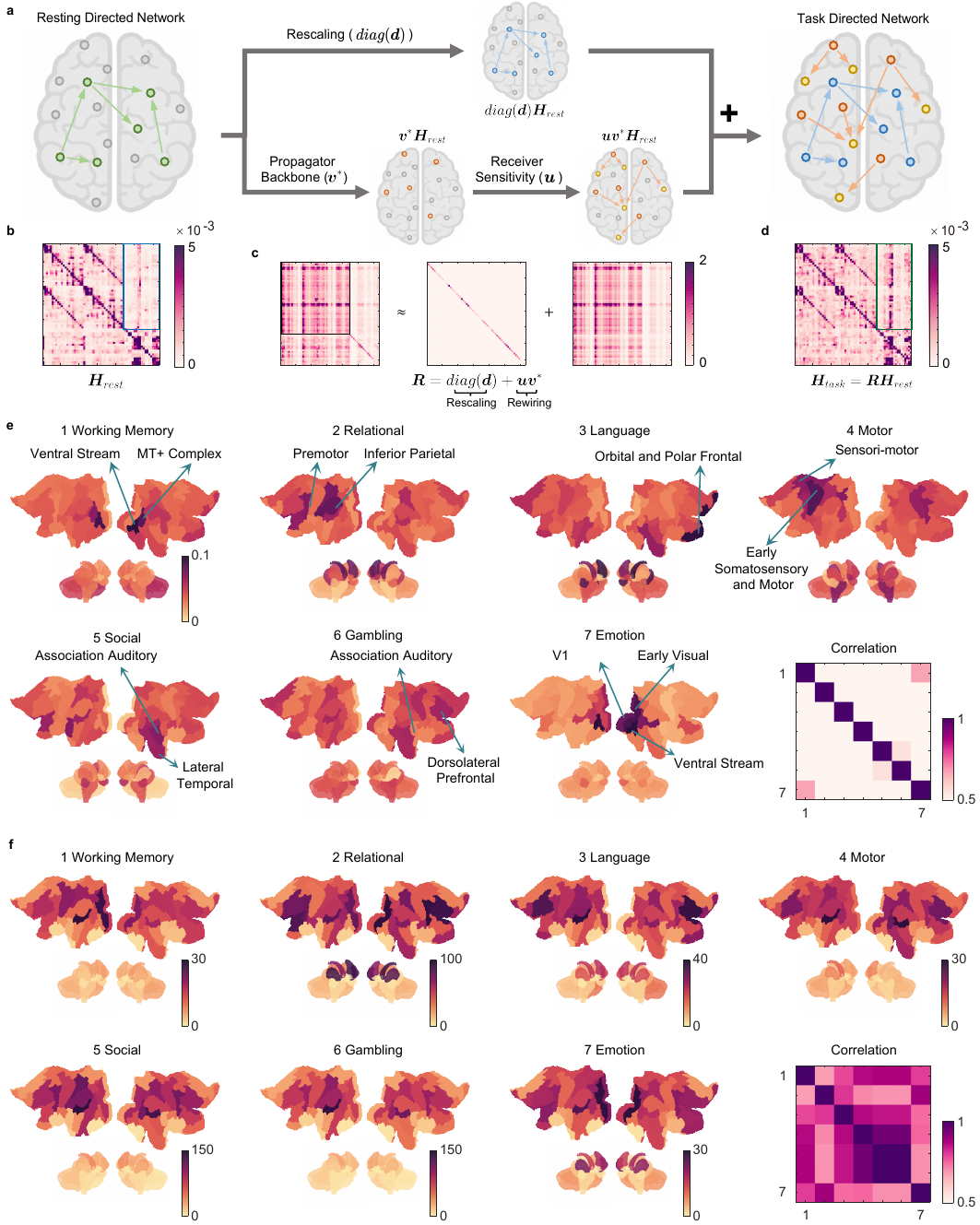} %
    	\caption{Task-induced reorganization of whole-brain functional coupling. $\bm{\mathrm{a}}$, The schematic of task-induced coupling modulation. $\bm{\mathrm{b,d}}$, The subcortical-to-cortical coupling has been enhanced during task states. $\bm{\mathrm{c}}$, Matrix decomposition reveals the low-rank structure of the relative coupling matrix $\bm{R} \approx \mathrm{Diag}(\bm{d}) + \bm{uv}^{*}$. Hence, subcortical-to-cortical enhancement $\bm{B}$ (green box) can be approximated by $\bm{CA}$ ($\bm{C}$: black box, $\bm{A}$: blue box). $\bm{\mathrm{e}}$, The factor $\bm{u}$ identifies task-active regions. $\bm{\mathrm{f}}$, The factor $\bm{v}$ shows cross-task consistency, as correlations among the 7 maps are greater than 0.7 ($p<1e-16$). Each map includes 2D flatmaps of the cerebral cortex along with bilateral views of subcortical structures.} 
        \label{fig:task}
\end{figure*}

We also apply our nonlinear model to task fMRI data and reveal a novel mechanism underlying directed network reconfiguration from the resting state to task engagement. Traditional analysis methods to assess changes in brain connectivity associated with tasks often rely on statistical models or signal processing techniques \cite{HUANG2024139}. These methods generally regard differences evoked by task stimuli as an additive effect on specific connections, based on the assumption that the influence of the task is local. For example, psychophysiological interaction (PPI) analysis identifies brain regions whose activity correlates more strongly with a seed region in a given psychological context (task condition) than in others \cite{Jill2012Tools}. However, task-evoked connectivity changes involve not only local connection adjustments but also dynamic reorganization of the global network. Hence, we hypothesize that these changes are multiplicative rather than additive.

To test this hypothesis, we calculate the coupling matrix $\bm{H}_t$ from different task-fMRI data in HCP \cite{2013Function} using the above data-driven paradigm. We discover a significant enhancement in subcortical-to-cortical coupling during task states (\Cref{fig:task}$\bm{\mathrm{b,d}}$, Extended Data \Cref{fig:task_appendix}$\bm{\mathrm{a}}$). Notably, we assess the task-rest difference in connectivity by a coupling modulation matrix $\bm{R} = \bm{H}_t \bm{H}_r^{-1}$, whose elements represent the global regulation of functional coupling between regions. The special structure of $\bm{R}$ allows it to be decomposed into a diagonal matrix $\mathrm{Diag}(\bm{d})$ and a rank-1 matrix $\bm{u} \bm{v}^{\rm{*}}$ (Sec. \ref{decom}; \Cref{fig:task}$\bm{\mathrm{c}}$, Extended Data \Cref{fig:task_appendix}$\bm{\mathrm{a}}$). Each element of $\bm{v}$, $v_i$, represents the weight or influence of the $i^{th}$ node as a propagator, while $u_i$ represents the gain or sensitivity of the $i^{th}$ node as a receiver. The local rescaling $d_i$ modulates the afferent coupling strength of the $i^{th}$ node, with effects predominantly manifested within subcortical regions (Extended Data \Cref{fig:task_appendix}$\bm{\mathrm{b-c}}$, Supplementary Information S11). The schematic of task-induced coupling modulation is presented in \Cref{fig:task}$\bm{\mathrm{a}}$.

The amplitude of $\bm{v}$ exhibits significant cross-task consistency (all task-pairs, $r>0.68, p \ll 0.001$, Pearson’s correlation; \Cref{fig:task}$\bm{\mathrm{f}}$). The regions with high amplitude involve the "task-positive" network, which activates during goal-directed, externally focused tasks requiring attention, problem-solving, and cognitive control \cite{bib6}. This result suggests a stable "backbone" for information propagation, acting as global hubs that relay information across the brain, regardless of task demands. In contrast, the amplitude of $\bm{u}$ demonstrates a distinct pattern in the specific task (\Cref{fig:task}$\bm{\mathrm{e}}$), corresponding to brain networks intrinsically involved in task-relevant cognitive operations \cite{bib6,bib33}. During the working memory task, a higher amplitude is observed in the MT+ complex and the ventral visual stream. Moreover, compared to the 0-back condition, the 2-back task elicits higher amplitudes in the prefrontal and parietal cortices, which are considered core components of the working memory network (\Cref{fig:task_appendix}$\bm{\mathrm{d}}$). Relational processing engages the premotor cortex and the inferior parietal lobule, while language processing preferentially activates the orbital and polar cortical territories. The motor task's $\bm{u}$ exhibits peak amplitudes in sensorimotor regions, particularly within the primary somatosensory and motor cortices. The social task demonstrates higher amplitude patterns in the association auditory cortex and lateral temporal regions. Gambling-related processing engages the association auditory cortex and dorsolateral prefrontal cortex with enhanced activation levels. $\bm{u}$ of the emotion task exhibits peak amplitudes in visual processing networks, particularly within the primary visual, early visual, and ventral stream cortical cortices. This implies that $\bm{u}$ highlights regions with heightened receptive sensitivity tailored to task demands. In summary, these findings reveal that the human brain achieves task adaptability through a hybrid architecture: A stable propagation backbone ($\bm{v}$) ensures efficient global integration and dynamic receptive tuning ($\bm{u}$) enables localized, task-specific processing.

\section{Discussion}\label{sec3}
In this study, we propose a novel Schr\"{o}dinger-form data-driven framework to characterize the complex-valued analytic architecture of brain dynamics. Our results demonstrate that this complex-valued framework substantially outperforms traditional real-valued models, achieving superior fit accuracy while generating more biologically plausible network structures and activity patterns. When applied to large-scale fMRI data, the framework uncovers significant age-related changes in whole-brain directed network structure and reveals a novel mechanism underlying task-induced coupling reconfiguration. These findings challenge the classical computational modeling paradigm and provide new insights into the functional organization of brain network dynamics.

The theoretical foundation of our approach is based on Hamiltonian mechanics, extending beyond previous applications that primarily focused on probabilistic inference models \cite{2016TheHamiltonianBrain,SENGUPTA20161107,baldy2025dynamic}. While earlier studies have employed Hamiltonian Monte Carlo for neural population modeling \cite{2016TheHamiltonianBrain} or unified mechanical and electrochemical behaviors within variational frameworks \cite{2015electromechanical}, these approaches fundamentally differ from ours in their treatment of neural states. Rather than mapping neuronal dynamics to the evolution of a Hamiltonian system, we treat observed neural states as generalized coordinates and introduce auxiliary signals as their conjugate momenta, thus establishing a complete whole-brain Hamiltonian system \cite{1983Mathematical,2015An}. Our results demonstrate that the auxiliary signals converge to the Hilbert transform of the observed signals \cite{2023Different,Horel1984ComplexPCA,2014Survey}. This mathematical relationship naturally gives rise to analytic signals governed by a linear Schr\"{o}dinger-like equation, thus providing an elegant and tractable framework for modeling whole-brain dynamics in the complex-valued field.

The Schr\"{o}dinger-like equation has emerged as a powerful tool to understand key brain dynamics \cite{bib32,bib81,Deco2025}, and our framework significantly advances this research direction by demonstrating its effectiveness in characterizing large-scale spatiotemporal brain organization. It achieves high accuracy in predicting high-resolution whole-brain recordings over short time, outperforming traditional models by addressing precision limitations associated with parameter redundancy \cite{bib16,bib30}. Through simulation of stimulus-evoked dynamics, the linear Schr\"{o}dinger-like framework reveals hierarchical timescales of cortical information processing that correspond closely to established neuroscientific principles \cite{bib66,bib67,bib68}, providing convergent validation across multiple levels of brain organization. In characterizing resting-state dynamics, our approach captures essential spatiotemporal modes that align well with functional networks \cite{bib27,bib28}. Unlike traditional complex PCA methods that primarily characterize signal propagation and inter-regional correlations \cite{bib44}, our framework directly identifies the dynamic features and temporal properties of spatiotemporal modes. The biological relevance of our complex-valued coupling is further validated by significantly higher SFC values compared to conventional functional connectivity measures \cite{bib37,bib38,bib39,bib40}. These findings highlight the importance of complex-valued analytic frameworks in neuroscience, which offer a powerful approach to elucidate the mechanisms underlying brain dynamics and complex cognitive processes.

Beyond linear Hamiltonian systems, we further construct a nonlinear data-driven paradigm to reveal non-equilibrium characteristics of brain dynamics. While existing complex-valued methods have independently modeled
the interregional connectivity of the real and imaginary components of signals \cite{bib3,bib49,bib47} and primarily focus on generating synthetic fMRI data \cite{2024Human,bib80}, our approach fundamentally differs by employing unified complex-valued interregional coupling to directly model analytic fMRI signals. This data-driven framework also results in a nonlinear Schr\"{o}dinger-like governing equation on brain graphs. Multifaceted assessments have shown that the complex-valued nonlinear model outperforms its real-valued counterpart in data fitting, biological interpretability, and signal transmission \cite{effec_connec}. Critically, our framework captures brain asymmetry through the complex-valued asymmetric coupling matrix, which exhibits strong correlations with the Granger causality measures. This asymmetric coupling architecture provides direct quantitative access to directional information flow \cite{2022Asymmetric} and temporal irreversibility \cite{2022Temporal,2024Broken}, which are the fundamental signatures of non-equilibrium neural dynamics \cite{2024Mind,Nartall,2025Non}. Taken together, these insights demonstrate that the nonlinear Schr\"{o}dinger-like model comprehensively reveals the fundamental principles governing the whole-brain non-equilibrium dynamics, providing an innovative perspective for understanding complex functions of healthy brains.

Another notable finding is that the estimated complex-valued coupling adapts throughout the lifespan. Unlike conventional FC methods for calculating lifespan growth curves \cite{sun2025lifespan}, our approach can capture more significant age-related changes in whole-brain functional coupling. With advancing age, we observe a progressive decrease in the coupling sparsity and mean coupling strength. For a larger spatial scale, these coupling indicators tend to plateau in later life, potentially suggesting the presence of distinct aging subtypes \cite{Subtypes}. This result resembles the patterns of developmental synaptic pruning documented in the human brain \cite{bib69, bib70} and aligns with age-related fluctuations in blood perfusion \cite{bib73}. Such multimodal indicators could concurrently reflect underlying neurodevelopmental and aging processes, providing a critical benchmark for quantifying individual differences in development, aging, and neuropsychiatric disorders \cite{sun2025lifespan,2023Publisher,0The}. 

Our study further elucidates the regulatory mechanisms of whole-brain coupling during task execution. Through analysis of task-rest relative coupling, we demonstrate that task stimuli simultaneously engage two distinct processes: local regulation of resting-state coupling through a scaling factor ($\bm{d}$), and global reconfiguration of resting-state dynamics via a novel low-rank structure decomposed into task specificity ($\bm{u}$) and task consistency ($\bm{v}$) components. The dual-process architecture fundamentally challenges classical neural models which assume a linear additive relationship between task and resting-state dynamics \cite{HUANG2024139,Jill2012Tools}. It provides a unified framework for understanding how task stimuli exert their influence on resting-state dynamics, offering new insights into the neural basis of cognitive flexibility and task control.

Together, these findings position Schr\"{o}dinger-like equation in the complex-valued field as a powerful, generalizable paradigm for analyzing functional neuroimaging data, one that promises new biomarkers for neuroscience. Meanwhile, our work offers a fresh perspective on modeling neural dynamics, clarifying how richer latent dynamics can be inferred from limited observations.

\section{Methods}\label{sec4}

\subsection{HCP data}
We analyze high-quality fMRI data from three Human Connectome Project (HCP) cohorts: HCP Young Adults (HCP-YA)\cite{bib43}, Development (HCP-D)\cite{bib72}, and Aging (HCP-A)\cite{bib71}.  The combined sample comprises 1,321 unrelated, healthy participants: young adults aged 22–37 years, adolescents under 22, and middle-aged to elderly individuals aged 40–100. Each participant underwent resting-state fMRI over two consecutive days, with two 15-minute runs per session (4 runs total), collected at 2mm isotropic resolution, TR=720ms, consistent with HCP acquisition protocols. For HCP-YA subjects, task-based fMRI followed the same acquisition scheme. All data is preprocessed using the HCP minimal pipeline (FSL, FreeSurfer, FIX-based ICA denoising), including correction for spatial and gradient distortions, motion, registration to T1-weighted structural scans, and normalization to standard space. For details on acquisition and preprocessing procedures, see Glasser et al.\cite{bib43}.

\subsection{fMRI data pre-processing}

For subsequent analyses, we impose an additional post-processing layer on the minimally pre-processed fMRI data. Vertex-wise BOLD time courses are first standardized (z-scored) across time to remove mean-variance differences. We then aggregate the signals to the regional scale by averaging time series within each anatomical label of a composite parcellation: the Human Connectome Project multimodal atlas template (HCP-MMP v1.0) \cite{huang2022extended}, subcortical voxels the Brainnectome atlas \cite{2016The}, and cerebellar/brainstem voxels the Shen268 atlas \cite{2013Groupwise}.  For analyses retaining voxel/vertex resolution, greyordinate data ($\sim$96 k) is resampled to a coarser grid. Cortical data is projected onto a 4 k-density fsLR surface template, whereas subcortical volumes are down-sampled within MNI space, thereby preserving anatomical correspondence across spaces.

In our study, we directly apply the Hilbert transform to our preprocessed BOLD signals to characterize neural dynamics, yielding complex-valued analytic signals that retain the real component while encoding instantaneous phase and amplitude information \cite{bib41,Horel1984ComplexPCA}. This transform effectively removes negative-frequency components in the spectral domain, providing simultaneous access to both amplitude and phase metrics and enabling detection of neural oscillatory patterns \cite{2019Cycle,0Time}. 

\subsection{Whole-brain Hamiltonian framework}\label{Hilbert}
Hamiltonian mechanics provides the framework that any conservative system can be described through a pair of conjugate generalized coordinates $\bm{q}$ and momenta $\bm{p}$ \cite{2015An,2016TheHamiltonianBrain}. This pair of variables is governed by Hamilton's equations \eqref{Hamilton1}, where $H$ is the Hamiltonian of the dynamical system. 
\begin{equation}\label{Hamilton1}
    \frac{{\rm d} \bm{p}}{{\rm d}t}= -\frac{\partial{H}}{\partial{ \bm{q}}}, \quad\frac{{\rm d} \bm{q}}{{\rm d}t}= \frac{\partial{H}}{\partial{ \bm{p}}}.
\end{equation}
In fact, the Hamiltonian is constituted by the potential energy $V(\bm{q})$ and the kinetic energy $T(\bm{p})$, resulting in the expression $ H(\bm{q},\bm{p}) =  V(\bm{q}) + T(\bm{p})$. Our framework utilizes a quadratic Hamiltonian, $H(\bm{q},\bm{p}) = -\big(\bm{q}^{\top} \bm{H} \bm{q} + \bm{p}^{\top} \bm{H} \bm{p} \big)/2$, derived from the quadratic potential and kinetic energies, $V(\bm{q}) = -\bm{q}^{\top} \bm{H} \bm{q}/2$ and $T(\bm{p}) = -\bm{p}^{\top} \bm{H} \bm{p}/2$, respectively. Analytical derivations have revealed that the conjugate momenta $\bm{p}$ are demonstrably the Hilbert transform of the generalized coordinates $\bm{q}$ when the coupling parameter $\bm{H}  = (h_{jk}) \in\mathbb{R}^{N \times N}\succ 0$ (a detailed proof is provided in Supplementary Information S1). This finding provides a novel perspective on understanding the interrelationship between these dual variables and strongly supports the use of the Hilbert transform to augment the measured dynamics with latent degrees of freedom \cite{2023Different}.

In our work, we regard the observed signals as generalized coordinates along with their conjugate auxiliary signals as generalized momenta. Since this system of coupled harmonic oscillators follows Hamilton's equations \eqref{2.1-1}, we develop a model to estimate the coupling parameter $\bm{H}$ and auxiliary signals $\bm{p}$ from the observed signals $\bm{q}$ (\Cref{fig:Linear}$\bm{\mathrm{a}}$). The neural network $F[\cdot]$ employs an architecture consisting of convolutional layers, fully connected layers, and additional convolutional layers,  establishing a correspondence between each node-wise auxiliary signal and its conjugate coordinate. The linear layer $\bm{H}$ is assumed to be symmetric with positive diagonals. The loss function is defined as equation \eqref{Loss}. Numerical experiments also reveal that the learned auxiliary signals correspond to the Hilbert transform of observed signals, indicating a necessary condition for characterizing the brain's dynamical system as a Hamiltonian system.

\begin{equation}\label{Loss}
    \mathcal{L} =  \sum_{j=1}^{N} \sum_{t=1}^{T} \Bigg(\frac{ {\rm d} \,q_j(t)}{{\rm d} t}  +  \sum_{k=1}^{N}  h_{jk} \cdot F[q_j](t) \Bigg)^2 + \Bigg( \frac{ {\rm d}F[q_j](t) }{{\rm d} t}  - \sum_{k=1}^{N}h_{jk} \cdot q_k(t)\Bigg)^2. 
\end{equation}

\subsection{Modeling resting-state dynamics using a linear Schr\"{o}dinger-like equation}\label{Group_linear}
Building upon the theoretical framework described above, we construct a data-driven model based on a linear Schr\"{o}dinger-like equation that directly uses analytic signals to derive the symmetric coupling matrix $\bm{H}$. For undirected inter-node coupling, the symmetric matrix $\bm{H}$ ensures that $\bm{Q}=\exp({\bf i}\bm{H} \delta t) $ remains unitary ($\delta t$ = 1 TR, TR = 720 ms). The linear Schr\"{o}dinger-like equation admits explicit-form solutions:
\begin{equation}\label{linear_solution}
\begin{split}
    \bm{\psi} (t + \delta t) &= \exp({\bf i}\bm{H}\delta t) \cdot \bm{\psi}(t) 
    = \bm{Q} \, \bm{\psi}(t),  \quad \bm{Q^*Q} = \bm{I},\\
\end{split}
\end{equation}
where the unitary evolution operator $\bm{Q}$ governs the temporal dynamics of the system with 1 TR duration and $\bm{I}$ is the identity matrix \cite{bib27}. This framework \eqref{linear_solution} enables the reconstruction and prediction of voxel-wise rs-fMRI signals, with parameter $\bm{Q}$ estimated by solving a unitary Procrustes problem ($S$ = 1, $T$ = 300 TRs in equation \eqref{8}; see Supplementary Information S2 for numerical implementation). Reconstruction and prediction accuracy are evaluated using Pearson’s correlation between the real component of the simulated time series and the empirical fMRI data (Supplementary Information S3).

To identify unified spatiotemporal characteristics of brain dynamics, we estimate the group-level transfer matrix $\bm{Q}$ by solving the unitary Procrustes problem \eqref{8} using voxel-level resting-state analytic signals spanning 1000 TRs from 1,321 participants:
\begin{equation}
\begin{aligned} \label{8}
\min_{\bm{Q^*Q} = \bm{I}} \quad & \sum_{s=1}^S \sum_{t=1}^T \Big\| \bm{Q\psi}_t^{(s)} - \bm{\psi}_{t+1}^{(s)}   \Big\|_2^2 ,\\   
\end{aligned}
\end{equation}
where $ \bm{\psi}_{t}^{(s)}$ denotes the state at time $t$ of the participant $s$. Further details on solving this problem are provided in Supplementary Information S2.

\subsection{Quantifying intrinsic timescales of cerebral cortex}\label{timescale}
We evaluate the performance of the linear Schr\"{o}dinger-like model in capturing the propagation characteristics of stimulus-evoked neural responses. Given the group-level transfer matrix $\bm{Q}$ within one TR and its eigenvalue decomposition $\bm{Q} = \bm{U}\bm{\Sigma}\bm{U}^*$, we define the unitary matrix $\bm{Q}_t = \bm{Q}^{\frac{1}{100}} = \bm{U}\bm{\Sigma}^{\frac{1}{100}} \bm{U}^*$, where $\bm{\Sigma} = \rm{diag}(\sigma_1,\sigma_2,...,\sigma_N)$, to govern the temporal evolution of the dynamics evoked by the stimulus over a time interval of 0.01 TR. Given that the expression $\sigma^{\frac{1}{100}}$ yields 100 complex roots, we select the root with the smallest positive frequency to construct $\bm{Q}_t$, following the criterion of maximum stability. An initial localized stimulus of amplitude 100 is delivered to all voxels within the left primary visual cortex. The experiment is then simulated with temporal evolution \eqref{linear_solution} governed by $\bm{Q}_t$ within 20 TRs. Based on the response signals in other voxels, we define intrinsic timescales to quantify information processing ability.

Previous research has quantified intrinsic timescales to identify the signal processing mechanisms of brain dynamics after V1 stimulation, which are determined by the decay time of the autocorrelation function of neural activity fluctuations \cite{bib74,bib75}. Some studies fit the autocorrelation function to an exponential model and define intrinsic timescales using the inherent parameters of this model \cite{bib66,bib67}. Due to the oscillatory properties of the autocorrelation function in our model (\Cref{fig:Linear}$\bm{\mathrm{g}}$), fitting the function using an exponential model is not advisable. Hence, we quantify the intrinsic timescale as the duration required for the autocorrelation to decay to 0.6. Regional intrinsic timescales are calculated by averaging voxel-wise timescales within each brain region. Subsequently, we have analyzed whether the intrinsic timescales of the cerebral cortex align with the hierarchical organization of human visual processing, particularly across two distinct visual pathways: the ventral visual pathway (V1, V2, V3, V4, PIT, TE, TGv/STSv, OFC), which is associated with object recognition and form processing, and the dorsal visual pathway (V1, V2, V3, V3A, MT, MST, LIP, ACC, 8Ad/46), which is specialized for spatial awareness and motion perception \cite{bib64,bib65}.

\subsection{Spatiotemporal mode decomposition of brain activity}\label{spatio}
The spatiotemporal modes of brain dynamics are derived from the eigenvalue decomposition of the group-level transfer matrix $\bm{Q} =\sum_k \sigma_k \bm{u}_k \bm{u}_k^*$, where each $\bm{u}_k$ represents a spatiotemporal mode. Since $\bm{Q}$ is unitary, its eigenvalues lie in the unit circle as $\sigma_k = e^{{\bf i}\lambda_k t}$, with all frequencies positive (\Cref{fig:mode}$\bm{\mathrm{b}}$). The spatiotemporal modes are then ordered by these oscillatory frequencies $\lambda_k$, forming an ascending sequence $0\le \lambda_1 \le \lambda_2 \le \dots$, where each frequency $\lambda_k$ represents the $k^{th}$ eigenvalue of the real-valued coupling matrix $\bm{H}$. Therefore, resting-state dynamics in the complex-valued field can be expressed as a weighted sum of the spatiotemporal modes:
\begin{equation}
     \bm{\psi}(t) = \bm{Q} \cdot \bm{\psi}(0) = \sum_k \sigma_k \bm{u}_k \big(\bm{u}_k^*\bm{\psi}(0)\big) = \sum_k \zeta_k \,e^{{\bf i} \lambda_k t} \bm{u}_k. \label{9}
\end{equation}
Each eigentriplet $\{ \bm{u}_k,\sigma_k, \zeta_k\}$ describes an evolutionary principle of the dynamical system with the mode $\bm{u}_k$, the oscillatory frequency $\lambda_k$ and the amplitude $\zeta_k = \bm{u}_k^*\bm{\psi}(0)$. Note that the first frequency $\lambda_1$ is approximately zero, and the corresponding spatiotemporal mode $\bm{u}_1$ exhibits the slowest temporal dynamics, as described by $\bm{\psi}_1 (t) =  e^{\mathrm{i} \lambda_1 t} \bm{u}_1$. Since the amplitude $\zeta_k$, which quantifies the participation level of the $k^{\mathrm{th}}$ spatiotemporal mode across different trajectories, is not time-varying, it can be disregarded in our analysis. Therefore, we focus on the evolution of the three most prominent low-frequency spatiotemporal modes (\Cref{fig:mode}).

\subsection{Derivation of structural eigenmodes}\label{structural_eigenmodes}
Structural eigenmodes are derived according to an eigenvalue decomposition of the structural Laplacian matrix, which is usually used to analyze the dissociation of various diffusion processes in previous studies \cite{bib34,bib35,bib45}. The structural connectivity matrix $\bm{C}$ represents an undirected weighted adjacency matrix of the brain network. The structural Laplacian matrix is defined as $\bm{L} = \bm{D} - \bm{C}$, where $\bm{D}$ is the diagonal degree matrix. To eliminate the influence of network size and density, the structural Laplacian matrix is normalized as $\bm{L}_0 = \bm{L} / \nu_{max}$ \cite{bib36}, where $\nu_{max}$ is the maximum eigenvalue of $\bm{L}$. By performing eigenvalue decomposition of the normalized structural Laplacian, $\bm{L}_0 = \sum_{k} \nu_k \bm{x}_k \bm{x}_k^{\rm{T}}$, we define each resulting eigenvector $\bm{x}_k$ as a structural eigenmode with spatial frequency $\nu_k$. Since the Laplacian matrix $\bm{L}_0$ is symmetric and positive definite, its eigenvalues are non-negative real numbers. These eigenvalues are ordered sequentially according to the spatial frequency of each eigenmode, that is, $0\le \nu_1 \le \nu_2 \le \dots$. The space spanned by the structural eigenmodes has variable dimensionality depending on the number of modes included. We introduce these eigenmodes for direct comparison with our spatiotemporal modes when reconstructing rs-fMRI (\Cref{fig:mode}$\bm{\mathbf{c}}$, Sec. \ref{sec_FC_recon}).

\subsection{Resting-state fMRI reconstruction using dynamics modes and structural eigenmodes}\label{sec_FC_recon}
We reconstruct resting-state dynamics within constrained subspaces using fixed sets of spatiotemporal modes, with mode weights estimated through least squares. We sequentially select the $k$ lowest frequency dynamic spatiotemporal modes $\bm{U}_k = [\bm{u}_1, \bm{u}_2, \ldots, \bm{u}_k]$ and compute the reconstructed signals as $\hat{\bm{\psi}} = \bm{U}_k \, \bm{U}_k^{\rm{*}} \, \bm{\psi}$, where $\bm{U}_k^{\rm{*}} \, \bm{\psi}$ represents the optimal weights of these $k$ spatiotemporal modes. This approach determines whether analytic signals lie within selected mode subspaces, assessing their low-frequency eigenspace representation. Since spatiotemporal modes yield complex-valued analytic signals, we use only real components to enable direct comparison with real-valued structural eigenmodes. Reconstruction accuracy is quantified using Pearson's correlation between empirical and reconstructed functional connectivity matrices. For structural eigenmode reconstruction, we apply the same principle to real-valued signals $\hat{\bm{q}} = \bm{X}_k \, \bm{X}_k^{\rm{T}} \, \bm{q}$, where $\bm{X}_k = [\bm{x}_1, \bm{x}_2, \ldots, \bm{x}_k]$ contains the $k$ lowest spatial frequency structural eigenmodes. The reconstructed accuracy is computed for subspaces generated by an increasing number of modes ($k=1,2,...,d$). A comparison of the accuracy curves for low-frequency spatiotemporal modes and structural eigenmodes (\Cref{fig:mode}$\bm{\mathrm{c}}$) reveals a significant difference in their data reconstruction capacity across matched mode numbers.

\subsection{Data-driven nonlinear modeling of brain dynamics using analytic signals}\label{data-driven}
Inspired by the above theoretical and experimental insights, we develop a unified nonlinear framework for modeling complex-valued analytic signals. Our complex-valued parametric framework combines nonlinear self-coupling dynamics with linear interregional coupling. This yields the unified model:
\begin{equation}
    \frac{{\rm d}\,\bm{\psi}}{{\rm d}\,t} = p(\bm{\psi}) + \bm{W} \bm{\psi},\label{1}
\end{equation}
where $\bm{\psi} \in \mathbb{C}^N$ represents the analytic signals, governed by a polynomial function $p(\bm{\psi})$ capturing the nonlinear self-dynamics within each region, and a coupling matrix $\bm{W} = (w_{jk}) \in \mathbb{C}^{N \times N}$ describing the linear interregional connectivity.

By replacing the symmetry constraint on the coupling matrix with a sparsity constraint, the whole-brain model more closely aligns with sparsity theories based on neural activity and synaptic connectivity in the human brain \cite{bib69,bib70}. The sparsity of the matrix $\bm{W}$ is controlled by a parameter $\mu$, which is determined via a cross-validation method (Extended Data \Cref{fig:7}$\bm{\mathrm{b-c}}$, Supplementary Information S7). Moreover, $p(\bm{\psi})$ is specified as a cubic polynomial: $p(\bm{\psi}) = \theta^{(1)} + \theta^{(2)} \bm{\psi} +\theta^{(3)} \bar{\bm{\psi}} + \dots+ \theta^{(10)} \bar{\bm{\psi}}^3$, where $\bm{\theta} \in \mathbb{C}^{10}$ represents optimized parameters that govern nonlinear nodal dynamics. The derivative $\rm{d} \bm{\psi} / \rm{d} t$ is numerically approximated by a discretization $D(\bm{\psi})$, the fourth-order finite-difference method. The model parameters, comprising the polynomial coefficients $\bm{\theta}$ and the sparse complex-valued coupling matrix $\bm{W}$, are estimated by solving the sparse matrix optimization problem \eqref{A} via the alternating direction method of multipliers (Extended Data \Cref{fig:7}$\bm{\mathrm{a}}$). Further algorithmic details are provided in Supplementary Information S6.

\begin{equation}
\begin{aligned} \label{A}
\min_{\bm{\theta}, \bm{W} } \quad & \frac{1}{2} \sum_{t = 1}^{\top} \sum_{j = 1}^N \Bigg\| p\big(\psi_j(t)\big) + \sum_{k=1}^N w_{jk} \psi_k(t) -D \big(\psi_j(t)\big) \Bigg\|_F^2 + \mu\|\bm{W}\|_1 .\\
\end{aligned}
\end{equation}

The nonlinear real-valued framework models the observed signals using the coupling matrix $\bm{W} \in \mathbb{R}^{N \times N}$ and the polynomial function $p(\bm{q}) = \theta^{(1)} + \theta^{(2)} \bm{q} + \theta^{(3)} \bm{q}^2 + \theta^{(4)} \bm{q}^3$ with parameters $\boldsymbol{\theta} \in \mathbb{R}^4$. These parameters are estimated from fMRI data using sparse regression.

\subsection{Structure-informed nonlinear model}\label{SC_constraint}
Considering a more biologically plausible framework, we use a traditional nonlinear complex-valued model (hopf \cite{bib3,bib49}) to simulate specific brain dynamics, in which node dynamics is coupled via an empirical structural connectivity. The temporal activity $\psi_j$ of each brain region $j$ is defined by the general equation \eqref{hopf}:
\begin{equation}
    \frac{{\rm d}\,\psi_j}{{\rm d}\,t} = p(\psi_j) + g\sum_{k=1}^N \bm{C}_{jk} \psi_k,\label{hopf}
\end{equation}
where $\bm{C}_{jk}$ represents the structural connectivity between nodes $\psi_j$, and $\psi_k$ and $g$ is a global coupling parameter that scales the connectivity between regions. In our study, we propose this structure-informed model as a fitting framework for analytic signals. The nonlinear function $p(\cdot)$  maintains the same mathematical form as in our unified nonlinear framework. To this end, we estimate the full set of model parameters, including the polynomial coefficients $\boldsymbol{\theta} \in \mathbb{C}^{10}$ and the global parameter $g \in \mathbb{C}$, by minimizing the least-squares error. A comprehensive analysis of the phase distribution of the resulting global parameter $g$ is also conducted. The analogous structure-informed model is applied to fit the real-valued signals $\bm{q}$, using the real-valued parameters $\boldsymbol{\theta} \in \mathbb{R}^4$ and $g \in \mathbb{R}$. These models enable simulation of stimulus-evoked dynamics under structural connectivity constraints in both complex-valued and real-valued fields.

\subsection{Modeling stimulus-evoked dynamics under structural connectivity constraint}\label{stimulus-evoked}
We have quantitatively evaluated the effectiveness of structure-informed nonlinear models (Sec. \ref{SC_constraint}) in capturing evoked neural responses across the cerebral cortex. To investigate localized cortical responses, we deliver an instantaneous stimulus (amplitude = 10) to primary sensory regions, including visual, auditory, motor, and somatosensory cortices, as defined by the HCP-MMP atlas. Subsequently, we simulate neural dynamics for a duration of 10 TRs (TR = 720 ms) with a temporal resolution of 0.01 TR, using structurally constrained models (both complex- and real-valued; Sec. \ref{SC_constraint}). The complete signal propagation process is illustrated in Supplementary Video S13.

To quantify stimulus-evoked information transmission, we characterize regional activity by the timing and amplitude of the initial response peak, which index the transmission rate and efficiency, respectively. We compute these features in both real-valued and complex-valued frameworks to assess their effectiveness in capturing stimulus-evoked responses under structural connectivity constraints (\Cref{fig:Nonlinear}$\bm{\mathbf{h}}$). The propagation speed and efficiency of various brain regions correspond to their communicability (Sec. \ref{network}) to the stimulated region.

\subsection{Network measures of brain coupling}\label{network}

Based on the existing literature, several measures are commonly employed in brain network research to assess the functional and structural complexity within brain networks \cite{bib77,bib78}. We use these network measures to compare the coupling matrices derived from the real-valued and complex-valued models.

\begin{itemize}
    \item The network measure, communicability ($G$), is utilized to evaluate the efficiency of information transmission across various paths among all nodes in the network. The communicability between nodes $i$ and $j$ is defined as
\begin{equation}
    \bm{G}_{ij} = \Big(e^{\bm{D}^{-\frac{1}{2}}\bm{H}\bm{D}^{-\frac{1}{2}}}\Big)_{ij}.
\end{equation}
Note that the network connectivity matrix $\bm{H}$ is normalized by $\bm{D} = \mathrm{Diag}(d_i)$, where $d_i = \sum_{k=1}^N \bm{H}_{ik}$ is the generalized weighted in-degree of the node $i$. The mean communicability of the network is $G =  \langle\bm{G}_{ij}\rangle$. The communicability of node $i$ is the $i^{th}$ row of $\bm{G}$, which is used to quantify the structural pathways of all lengths to the $i$-th region, including shortest-path and higher-order connections that may be indirect between nodes.
\item The network density ($k$) reflects the overall weighted connectivity of a network. Denote $k= \frac{1}{N(N-1)}\sum_{ij} \bm{H}_{ij}$, where $N$ is the number of nodes. 
\item The modularity index ($Q$) quantifies the effectiveness of partitioning a network into distinct groups. A higher modularity of the coupling matrix indicates a greater ability to divide the functional regions of the brain network. The modularity index for the undirected network derived from the coupling matrix $\bm{H}$ is expressed by
\begin{equation}
    Q = \frac{1}{2 E} \sum_{i,j} \Bigg(\bm{H}_{ij}-\frac{d_id_j}{E} \Bigg) \delta(\bm{\psi}_i,\bm{\psi}_j),    
\end{equation}
where $\delta(\bm{\psi}_i,\bm{\psi}_j)$ is called the Kronecker delta function and equals one if $\bm{\psi}_i$ and $\bm{\psi}_j$ belong to the same functional region and zero otherwise. Let $E = \sum_{i,j} |\bm{H}_{ij}|$ represent the total coupling strength of the network. 
\item Assortativity ($r$) refers to the tendency of nodes in a network to connect with other similar nodes. A network with high assortativity typically features a dense cluster of highly interconnected central nodes. Assortativity of an undirected network is expressed by
\begin{equation}
    r = \frac{\frac{1}{E} \sum_{i,j} d_i d_j \bm{H}_{ij} - \big(\frac{1}{E} \sum_{i,j} \frac{(d_i + d_j)}{2} \bm{H}_{ij}\big)^2}{\frac{1}{E}\sum_{i,j}\frac{(d_i^2 +d_j^2)}{2}\bm{H}_{ij}-\big(\frac{1}{E} \sum_{i,j}\frac{(d_i + d_j)}{2} \bm{H}_{ij}\big)^2}.
\end{equation}
\end{itemize}

\subsection{Mathematical definition in algorithms and analysis}\label{2.5}
This section details the mathematical concepts of our model, algorithms, and data analysis. The Frobenius norm of a matrix, denoted $\|\cdot\|_F$, is defined as the square root of the sum of the squares of its elements and quantifies the overall magnitude of the matrix. The $L_1$ norm, $\|\cdot\|_1$, given by the sum of the absolute values of the elements, is utilized in our algorithm as a regularization term.

The Hamiltonian represents the total energy of a brain dynamical system, taking a standard quadratic form, $H(\bm{\psi}) = -\bm{\psi}^*\bm{H}\bm{\psi}$, for linear dynamics \cite{bib26}. For nonlinear dynamics governed by ${\bf i} ( \frac{{\rm d}}{{\rm d}t}+ \bm{H} )\bm{\psi} = q(\bm{\psi})$, we obtain the corresponding Hamiltonian by evaluating $\bm{\psi}^*{\bf i} \frac{{\rm d} \bm{\psi}}{{\rm d}t}$, which leads to the following equation:
\begin{equation}
    H(\bm{\psi}) = \bm{\psi}^*{\bf i}  \frac{{\rm d} \bm{\psi}}{{\rm d}t} =\bm{\psi}^* \big(q(\bm{\psi}) - \bm{H} \bm{\psi}\big). \label{10}
\end{equation}
The corresponding total energy is defined as the temporal mean $\langle |H| \rangle_t$. Comprehensive analyses are detailed in Supplementary Information S10.

\subsection{The decomposition of the relative coupling matrix}\label{decom}

To quantify coupling differences between the resting state and the task conditions, we define the relative coupling matrix as $\bm{R} = \bm{H}{t} \bm{H}{r}^{-1}$, where $\bm{H}{t}$ and $\bm{H}{r}$ denote the group-averaged coupling matrices for the task states and the resting state, respectively. The task-state dynamics are then governed by:
\begin{equation}
    {\bf i} \frac{{\rm d}\,\bm{\psi}}{{\rm d}\,t} = q(\bm{\psi}) - \bm{H}_{t} \bm{\psi}=q(\bm{\psi}) - \bm{R}\, \bm{H}_{r} \bm{\psi}.
\end{equation}

Due to the special structure of the coupling modulation matrix $\bm{R}$ (\Cref{fig:task}$\bm{\mathbf{c}}$, Extended Data \ref{fig:task_appendix}$\bm{\mathbf{a}}$), we can decompose it as a superposition of a diagonal matrix $\mathrm{Diag}(\bm{d})$ and a rank-1 matrix $\bm{u}\bm{v}^*$ using an alternating descent method. Further details of the algorithm are provided in Supplementary Information S11. Consequently, the task-state coupling matrix $\bm{H}_t$ admits the following decomposition:

\begin{equation}
    \bm{H}_{t} = \bm{R}\bm{H}_{r} \approx \big( \mathrm{Diag}(\bm{d}) + \bm{u} \bm{v}^{\rm *}\big) \bm{H}_{r} = \mathrm{Diag}(\bm{d}) \bm{H}_{r}  + \bm{u} ( \bm{v}^{\rm *} \bm{H}_{r}).
\end{equation}
Note that the vectors $\bm{u}$ and $\bm{v}$ encode whole-brain coupling coefficients, representing global integration. In this framework, $v_i$ denotes the influence weight of the $i^{th}$ node in its role as a propagator, and $u_i$ denotes its sensitivity gain as a receiver. The diagonal matrix $\mathrm{Diag}(\bm{d})$ modulates the region-specific afferent coupling strengths during the resting state, which embodies a local scaling mechanism (Extended Data Fig. \Cref{fig:task_appendix}$\bm{\mathrm{b}}$). In summary, these factors collectively represent the primary determinants of task-induced dynamics, underscoring their pivotal role in mediating how task stimuli modulate resting-state brain activity (Extended Data \Cref{fig:task_appendix}$\bm{\mathrm{d}}$).

To evaluate the functional specificity of the cerebral cortex, we average the amplitudes of the diagonal scaling factor $\bm{d}$, the two vectors $\bm{u}, \bm{v}$, and the coupling matrices. This averaging is performed within 63 functionally defined regions of the HCP-MMP parcellation template \cite{bib6,huang2022extended}, which comprise the left hemisphere (regions 1–22), the right hemisphere (regions 23–44), and the subcortical areas (regions 45–63).

\section*{Data availability}
The Human Connectome project (HCP) dataset is publicly available at \href{http://www.humanconnectomeproject.org/data/}{http://www.humanconnectomeproject.org/data/}. The instructions for accessing HCP data can be found in \href{https://www.humanconnectome.org/}{https://www.humanconnectome.org/}.

\section*{Code availability}
The code for the main results of this paper is provided in \href{https://github.com/ShirleyZhang111/Schrodinger-Brain.git}{https://github.com/ShirleyZhang111/Schrodinger-Brain.git}.

\section*{Acknowledgements}

This work was partially supported by the National Natural Science Foundation of China (No. 12471481, U24A2001 to W.D.), the Science and Technology Commission of Shanghai Municipality (No. 23ZR1403000 to W.D.), and the Open Foundation of Key Laboratory Advanced Manufacturing for Optical Systems, CAS (No. KLMSKF202403 to W.D.), the Lingang Laboratory (No. LGL-1987 to W.L.), 111 Project (B18015 to J.F.) and Humboldt Research Award (to J.F.). G.D. is supported by Grant PID2022-136216NB-I00 funded by MICIU/AEI/10.13039/501100011033 and by “ERDF A way of making Europe”,  ERDF, EU, Project NEurological MEchanismS of Injury, and Sleep-like cellular dynamics (NEMESIS) (ref. 101071900) funded by the EU ERC Synergy Horizon Europe, and AGAUR research support grant (ref. 2021 SGR 00917) funded by the Department of Research and Universities of the Generalitat of Catalunya. The funders had no role in study design, data collection and analysis, the decision to publish, or the preparation of the manuscript.

\bibliography{sn-bibliography}

\begin{thebibliography}{10}
\expandafter\ifx\csname url\endcsname\relax
  \def\url#1{\burl{#1}}\fi
\expandafter\ifx\csname urlprefix\endcsname\relax\def\urlprefix{URL }\fi
\providecommand{\bibinfo}[2]{#2}
\providecommand{\eprint}[2][]{\url{#2}}
\providecommand{\doi}[1]{\url{https://doi.org/#1}}
\bibcommenthead

\bibitem{openstax2016motion}
\bibinfo{author}{{OpenStax}}.
\newblock \emph{\bibinfo{title}{Uniform Circular Motion and Simple Harmonic Motion}}  (\bibinfo{publisher}{Rice University}, \bibinfo{address}{Houston, TX}, \bibinfo{year}{2016}).

\bibitem{Leah2013The}
\bibinfo{author}{Portnow, L.~H.}, \bibinfo{author}{Vaillancourt, D.~E.} \& \bibinfo{author}{Okun, M.~S.}
\newblock \bibinfo{title}{The history of cerebral pet scanning: from physiology to cutting-edge technology}.
\newblock \emph{\bibinfo{journal}{Neurology}} \textbf{\bibinfo{volume}{80}}, \bibinfo{pages}{952--956} (\bibinfo{year}{2013}).

\bibitem{ZHANG2020116390}
\bibinfo{author}{Zhang, X.}, \bibinfo{author}{Pan, W.} \& \bibinfo{author}{Keilholz, S.~D.}
\newblock \bibinfo{title}{The relationship between bold and neural activity arises from temporally sparse events}.
\newblock \emph{\bibinfo{journal}{NeuroImage}} \textbf{\bibinfo{volume}{207}}, \bibinfo{pages}{116390} (\bibinfo{year}{2020}).

\bibitem{bib3}
\bibinfo{author}{Ponce-Alvarez, A.} \& \bibinfo{author}{Deco, G.}
\newblock \bibinfo{title}{The hopf whole-brain model and its linear approximation}.
\newblock \emph{\bibinfo{journal}{Scientific Reports}} \textbf{\bibinfo{volume}{14}}, \bibinfo{pages}{2615} (\bibinfo{year}{2024}).

\bibitem{bib22}
\bibinfo{author}{Deco, G.}, \bibinfo{author}{Kringelbach, M.~L.}, \bibinfo{author}{Jirsa, V.~K.} \& \bibinfo{author}{Ritter, P.}
\newblock \bibinfo{title}{The dynamics of resting fluctuations in the brain: metastability and its dynamical cortical core}.
\newblock \emph{\bibinfo{journal}{Scientific Reports}} \textbf{\bibinfo{volume}{7}}, \bibinfo{pages}{3095} (\bibinfo{year}{2017}).

\bibitem{bib47}
\bibinfo{author}{Deco, G.}, \bibinfo{author}{Jirsa, V.~K.} \& \bibinfo{author}{Mcintosh, A.~R.}
\newblock \bibinfo{title}{Emerging concepts for the dynamical organization of resting-state activity in the brain.}
\newblock \emph{\bibinfo{journal}{Nature Reviews Neuroscience}} \textbf{\bibinfo{volume}{12}}, \bibinfo{pages}{43--56} (\bibinfo{year}{2011}).

\bibitem{bib19}
\bibinfo{author}{Jobst, B.~M.}, \bibinfo{author}{Hindriks, R.},  \& \bibinfo{author}{Deco, G.}
\newblock \bibinfo{title}{Increased stability and breakdown of brain effective connectivity during slow-wave sleep: Mechanistic insights from whole-brain computational modelling}.
\newblock \emph{\bibinfo{journal}{Scientific Reports}} \textbf{\bibinfo{volume}{7}}, \bibinfo{pages}{4634} (\bibinfo{year}{2017}).

\bibitem{bib1}
\bibinfo{author}{Li, A.} \emph{et~al.}
\newblock \bibinfo{title}{Hierarchical fluctuation shapes a dynamic flow linked to states of consciousness}.
\newblock \emph{\bibinfo{journal}{Nature Communications}} \textbf{\bibinfo{volume}{14}}, \bibinfo{pages}{3238} (\bibinfo{year}{2023}).

\bibitem{bib21}
\bibinfo{author}{Kim, H.}, \bibinfo{author}{Moon, J.}, \bibinfo{author}{Mashour, G.~A.} \& \bibinfo{author}{Lee, U.}
\newblock \bibinfo{title}{Mechanisms of hysteresis in human brain networks during transitions of consciousness and unconsciousness: Theoretical principles and empirical evidence}.
\newblock \emph{\bibinfo{journal}{PLoS Computational Biology}} \textbf{\bibinfo{volume}{14}}, \bibinfo{pages}{e1006424} (\bibinfo{year}{2018}).

\bibitem{bib20}
\bibinfo{author}{Jobst, B.~M.} \emph{et~al.}
\newblock \bibinfo{title}{Increased sensitivity to strong perturbations in a whole-brain model of lsd}.
\newblock \emph{\bibinfo{journal}{NeuroImage}} \textbf{\bibinfo{volume}{230}}, \bibinfo{pages}{117809} (\bibinfo{year}{2021}).

\bibitem{bib56}
\bibinfo{author}{Patow, G.}, \bibinfo{author}{Martin, I.}, \bibinfo{author}{Sanz~Perl, Y.}, \bibinfo{author}{Kringelbach, M.~L.} \& \bibinfo{author}{Deco, G.}
\newblock \bibinfo{title}{Whole-brain modelling: an essential tool for understanding brain dynamics}.
\newblock \emph{\bibinfo{journal}{Nature Reviews Methods Primers}} \textbf{\bibinfo{volume}{4}}, \bibinfo{pages}{53} (\bibinfo{year}{2024}).

\bibitem{bib57}
\bibinfo{author}{Pathak, A.}, \bibinfo{author}{Roy, D.} \& \bibinfo{author}{Banerjee, A.}
\newblock \bibinfo{title}{Whole-brain network models: from physics to bedside}.
\newblock \emph{\bibinfo{journal}{Frontiers in Computational Neuroscience}} \textbf{\bibinfo{volume}{16}}, \bibinfo{pages}{866517} (\bibinfo{year}{2022}).

\bibitem{bib44}
\bibinfo{author}{Bolt, T.} \emph{et~al.}
\newblock \bibinfo{title}{A parsimonious description of global functional brain organization in three spatiotemporal patterns}.
\newblock \emph{\bibinfo{journal}{Nature Neuroscience}} \textbf{\bibinfo{volume}{25}}, \bibinfo{pages}{1093 -- 1103} (\bibinfo{year}{2022}).

\bibitem{bib32}
\bibinfo{author}{Deco, G.}, \bibinfo{author}{Sanz~Perl, Y.} \& \bibinfo{author}{Kringelbach, M.~L.}
\newblock \bibinfo{title}{Complex harmonics reveal low-dimensional manifolds of critical brain dynamics}.
\newblock \emph{\bibinfo{journal}{Physical Review E}} \textbf{\bibinfo{volume}{111}}, \bibinfo{pages}{014410} (\bibinfo{year}{2025}).

\bibitem{bib81}
\bibinfo{author}{Vohryzek, J.}, \bibinfo{author}{Sanz-Perl, Y.}, \bibinfo{author}{Kringelbach, M.~L.} \& \bibinfo{author}{Deco, G.}
\newblock \bibinfo{title}{Human brain dynamics are shaped by rare long-range connections over and above cortical geometry}.
\newblock \emph{\bibinfo{journal}{Proceedings of the National Academy of Sciences}} \textbf{\bibinfo{volume}{122}}, \bibinfo{pages}{e2415102122} (\bibinfo{year}{2025}).

\bibitem{2024Human}
\bibinfo{author}{Koller, D.~P.}, \bibinfo{author}{Schirner, M.} \& \bibinfo{author}{Ritter, P.}
\newblock \bibinfo{title}{Human connectome topology directs cortical traveling waves and shapes frequency gradients}.
\newblock \emph{\bibinfo{journal}{Nature Communications}} \textbf{\bibinfo{volume}{15}}, \bibinfo{pages}{3570} (\bibinfo{year}{2024}).

\bibitem{bib80}
\bibinfo{author}{Piccinini, J.} \& \bibinfo{author}{Deco, G.}
\newblock \bibinfo{title}{Data-driven discovery of canonical large-scale brain dynamics}.
\newblock \emph{\bibinfo{journal}{Cerebral Cortex Communications}} \textbf{\bibinfo{volume}{3}}, \bibinfo{pages}{tgac045} (\bibinfo{year}{2022}).

\bibitem{1983Mathematical}
\bibinfo{author}{Dirac, P.~A.}
\newblock \emph{\bibinfo{title}{The Lagrangian in quantum mechanics}}, \bibinfo{pages}{111--119} (\bibinfo{year}{2005}).

\bibitem{2015An}
\bibinfo{author}{Malham, S.~J.}
\newblock \bibinfo{title}{An introduction to lagrangian and hamiltonian mechanics}.
\newblock \emph{\bibinfo{journal}{Heriot-Watt University}} \textbf{\bibinfo{volume}{47}} (\bibinfo{year}{2016}).

\bibitem{hcp2017manual}
\bibinfo{author}{{Human Connectome Project Consortium}}.
\newblock \emph{\bibinfo{title}{HCP Young Adult: 900 Subjects Data Release — Reference Manual}}.
\newblock \bibinfo{organization}{Washington University–University of Minnesota} (\bibinfo{year}{2017}).

\bibitem{bib42}
\bibinfo{author}{Van~Essen, D.~C.} \emph{et~al.}
\newblock \bibinfo{title}{The wu-minn human connectome project: an overview}.
\newblock \emph{\bibinfo{journal}{NeuroImage}} \textbf{\bibinfo{volume}{80}}, \bibinfo{pages}{62--79} (\bibinfo{year}{2013}).

\bibitem{Horel1984ComplexPCA}
\bibinfo{author}{Horel, J.~D.}
\newblock \bibinfo{title}{Complex principal component analysis: Theory and examples}.
\newblock \emph{\bibinfo{journal}{Journal of Applied Meteorology and Climatology}} \textbf{\bibinfo{volume}{23}}, \bibinfo{pages}{1660--1673} (\bibinfo{year}{1984}).

\bibitem{2014Survey}
\bibinfo{author}{Singh, A.}
\newblock \bibinfo{title}{Survey paper on hilbert transform with its applications in signal processing}.
\newblock \emph{\bibinfo{journal}{Int J Comput Sci Inf Technol}} \textbf{\bibinfo{volume}{5}}, \bibinfo{pages}{3880--3882} (\bibinfo{year}{2014}).

\bibitem{2023Different}
\bibinfo{author}{Wodeyar, A.} \emph{et~al.}
\newblock \bibinfo{title}{Different methods to estimate the phase of neural rhythms agree but only during times of low uncertainty}.
\newblock \emph{\bibinfo{journal}{eNeuro}} \textbf{\bibinfo{volume}{10}}, \bibinfo{pages}{18} (\bibinfo{year}{2023}).

\bibitem{bib16}
\bibinfo{author}{Paula, S.~L.} \emph{et~al.}
\newblock \bibinfo{title}{The virtual brain: a simulator of primate brain network dynamics}.
\newblock \emph{\bibinfo{journal}{Frontiers in Neuroinformatics}} \textbf{\bibinfo{volume}{7}}, \bibinfo{pages}{10} (\bibinfo{year}{2013}).

\bibitem{bib51}
\bibinfo{author}{Breakspear, M.}
\newblock \bibinfo{title}{Dynamic models of large-scale brain activity}.
\newblock \emph{\bibinfo{journal}{Nature Neuroscience}} \textbf{\bibinfo{volume}{20}}, \bibinfo{pages}{340--352} (\bibinfo{year}{2017}).

\bibitem{2024Mind}
\bibinfo{author}{Kringelbach, M.~L.}, \bibinfo{author}{Perl, Y.~S.} \& \bibinfo{author}{Deco, G.}
\newblock \bibinfo{title}{The thermodynamics of mind}.
\newblock \emph{\bibinfo{journal}{Trends in Cognitive Sciences}} \textbf{\bibinfo{volume}{28}}, \bibinfo{pages}{14} (\bibinfo{year}{2024}).

\bibitem{deco2022insideout}
\bibinfo{author}{Deco, G.} \emph{et~al.}
\newblock \bibinfo{title}{The insideout framework provides precise signatures of the balance of intrinsic and extrinsic dynamics in brain states}.
\newblock \emph{\bibinfo{journal}{Communications Biology}} \textbf{\bibinfo{volume}{5}}, \bibinfo{pages}{572} (\bibinfo{year}{2022}).

\bibitem{bib66}
\bibinfo{author}{Murray, J.~D.} \emph{et~al.}
\newblock \bibinfo{title}{A hierarchy of intrinsic timescales across primate cortex}.
\newblock \emph{\bibinfo{journal}{Nature Neuroscience}} \textbf{\bibinfo{volume}{17}}, \bibinfo{pages}{1661--1663} (\bibinfo{year}{2014}).

\bibitem{bib67}
\bibinfo{author}{Li, S.} \& \bibinfo{author}{Wang, X.-J.}
\newblock \bibinfo{title}{Hierarchical timescales in the neocortex: Mathematical mechanism and biological insights}.
\newblock \emph{\bibinfo{journal}{Proceedings of the National Academy of Sciences}} \textbf{\bibinfo{volume}{119}}, \bibinfo{pages}{e2110274119} (\bibinfo{year}{2022}).

\bibitem{bib68}
\bibinfo{author}{Wolff, A.} \emph{et~al.}
\newblock \bibinfo{title}{Intrinsic neural timescales: temporal integration and segregation}.
\newblock \emph{\bibinfo{journal}{Trends in Cognitive Sciences}} \textbf{\bibinfo{volume}{26}}, \bibinfo{pages}{159--173} (\bibinfo{year}{2022}).

\bibitem{bib27}
\bibinfo{author}{Baddoo, P.~J.} \emph{et~al.}
\newblock \bibinfo{title}{Physics-informed dynamic mode decomposition}.
\newblock \emph{\bibinfo{journal}{Proceedings of the Royal Society A}} \textbf{\bibinfo{volume}{479}}, \bibinfo{pages}{20220576} (\bibinfo{year}{2023}).

\bibitem{bib34}
\bibinfo{author}{Pang, J.~C.} \emph{et~al.}
\newblock \bibinfo{title}{Geometric constraints on human brain function}.
\newblock \emph{\bibinfo{journal}{Nature}} \textbf{\bibinfo{volume}{618}}, \bibinfo{pages}{566--574} (\bibinfo{year}{2023}).

\bibitem{bib35}
\bibinfo{author}{Yang, Y.} \emph{et~al.}
\newblock \bibinfo{title}{Enhanced brain structure-function tethering in transmodal cortex revealed by high-frequency eigenmodes}.
\newblock \emph{\bibinfo{journal}{Nature Communications}} \textbf{\bibinfo{volume}{14}}, \bibinfo{pages}{6744} (\bibinfo{year}{2023}).

\bibitem{bib45}
\bibinfo{author}{Robinson, P.~A.} \emph{et~al.}
\newblock \bibinfo{title}{Eigenmodes of brain activity: Neural field theory predictions and comparison with experiment}.
\newblock \emph{\bibinfo{journal}{NeuroImage}} \textbf{\bibinfo{volume}{142}}, \bibinfo{pages}{79--98} (\bibinfo{year}{2016}).

\bibitem{bib46}
\bibinfo{author}{Gabay, N.~C.} \& \bibinfo{author}{Robinson, P.}
\newblock \bibinfo{title}{Cortical geometry as a determinant of brain activity eigenmodes: Neural field analysis}.
\newblock \emph{\bibinfo{journal}{Physical Review E}} \textbf{\bibinfo{volume}{96}}, \bibinfo{pages}{032413} (\bibinfo{year}{2017}).

\bibitem{bib28}
\bibinfo{author}{Thomas~Yeo, B.~T.} \emph{et~al.}
\newblock \bibinfo{title}{The organization of the human cerebral cortex estimated by intrinsic functional connectivity}.
\newblock \emph{\bibinfo{journal}{Journal of Neurophysiology}} \textbf{\bibinfo{volume}{106}}, \bibinfo{pages}{1125--1165} (\bibinfo{year}{2011}).

\bibitem{bib38}
\bibinfo{author}{Fotiadis, P.} \emph{et~al.}
\newblock \bibinfo{title}{Structure--function coupling in macroscale human brain networks}.
\newblock \emph{\bibinfo{journal}{Nature Reviews Neuroscience}} \textbf{\bibinfo{volume}{25}}, \bibinfo{pages}{688--704} (\bibinfo{year}{2024}).

\bibitem{bib37}
\bibinfo{author}{Baum, G.~L.} \emph{et~al.}
\newblock \bibinfo{title}{Development of structure--function coupling in human brain networks during youth}.
\newblock \emph{\bibinfo{journal}{Proceedings of the National Academy of Sciences}} \textbf{\bibinfo{volume}{117}}, \bibinfo{pages}{771--778} (\bibinfo{year}{2020}).

\bibitem{bib39}
\bibinfo{author}{Gu, Z.} \emph{et~al.}
\newblock \bibinfo{title}{Heritability and interindividual variability of regional structure-function coupling}.
\newblock \emph{\bibinfo{journal}{Nature Communications}} \textbf{\bibinfo{volume}{12}}, \bibinfo{pages}{4894} (\bibinfo{year}{2021}).

\bibitem{bib40}
\bibinfo{author}{Liégeois, R.} \emph{et~al.}
\newblock \bibinfo{title}{Revisiting correlation-based functional connectivity and its relationship with structural connectivity}.
\newblock \emph{\bibinfo{journal}{Network Neuroscience}} \textbf{\bibinfo{volume}{4}}, \bibinfo{pages}{1235--1251} (\bibinfo{year}{2020}).

\bibitem{bib69}
\bibinfo{author}{Kudithipudi, D.} \emph{et~al.}
\newblock \bibinfo{title}{Neuromorphic computing at scale}.
\newblock \emph{\bibinfo{journal}{Nature}} \textbf{\bibinfo{volume}{637}}, \bibinfo{pages}{801--812} (\bibinfo{year}{2025}).

\bibitem{2015UK}
\bibinfo{author}{Sudlow, C.} \emph{et~al.}
\newblock \bibinfo{title}{Uk biobank: an open access resource for identifying the causes of a wide range of complex diseases of middle and old age}.
\newblock \emph{\bibinfo{journal}{PLoS medicine}} \textbf{\bibinfo{volume}{12}}, \bibinfo{pages}{e1001779} (\bibinfo{year}{2015}).

\bibitem{nyc_taxi_2015}
\bibinfo{author}{{NYC Taxi and Limousine Commission}}.
\newblock \bibinfo{title}{Tlc trip record data} (\bibinfo{year}{2015}).
\newblock \urlprefix\url{https://www1.nyc.gov/site/tlc/about/tlc-trip-record-data.page}.

\bibitem{2013Nonequilibrium}
\bibinfo{author}{Yan, H.} \emph{et~al.}
\newblock \bibinfo{title}{Nonequilibrium landscape theory of neural networks}.
\newblock \emph{\bibinfo{journal}{Proceedings of the National Academy of Sciences}} \textbf{\bibinfo{volume}{110}}, \bibinfo{pages}{E4185--E4194} (\bibinfo{year}{2013}).

\bibitem{2022Temporal}
\bibinfo{author}{de~la Fuente, L.~A.} \emph{et~al.}
\newblock \bibinfo{title}{Temporal irreversibility of neural dynamics as a signature of consciousness}.
\newblock \emph{\bibinfo{journal}{Cerebral Cortex}} \textbf{\bibinfo{volume}{33}}, \bibinfo{pages}{1856--1865} (\bibinfo{year}{2023}).

\bibitem{2024Broken}
\bibinfo{author}{Nartallo-Kaluarachchi, R.} \emph{et~al.}
\newblock \bibinfo{title}{Broken detailed balance and entropy production in directed networks}.
\newblock \emph{\bibinfo{journal}{Physical Review E}} \bibinfo{pages}{110} (\bibinfo{year}{2024}).

\bibitem{bib79}
\bibinfo{author}{Deco, G.} \emph{et~al.}
\newblock \bibinfo{title}{Awakening: Predicting external stimulation to force transitions between different brain states}.
\newblock \emph{\bibinfo{journal}{Proceedings of the National Academy of Sciences}} \textbf{\bibinfo{volume}{116}}, \bibinfo{pages}{18088--18097} (\bibinfo{year}{2019}).

\bibitem{bib77}
\bibinfo{author}{Pigorsch, U.} \& \bibinfo{author}{Sabek, M.}
\newblock \bibinfo{title}{Assortative mixing in weighted directed networks}.
\newblock \emph{\bibinfo{journal}{Physica A: Statistical Mechanics and its Applications}} \textbf{\bibinfo{volume}{604}}, \bibinfo{pages}{127850} (\bibinfo{year}{2022}).

\bibitem{bib78}
\bibinfo{author}{Fornito, A.}, \bibinfo{author}{Zalesky, A.} \& \bibinfo{author}{Bullmore, E.}
\newblock \emph{\bibinfo{title}{Fundamentals of brain network analysis}}  (\bibinfo{publisher}{Academic press}, \bibinfo{year}{2016}).

\bibitem{bib64}
\bibinfo{author}{Kravitz, D.~J.} \emph{et~al.}
\newblock \bibinfo{title}{The ventral visual pathway: an expanded neural framework for the processing of object quality}.
\newblock \emph{\bibinfo{journal}{Trends in Cognitive Sciences}} \textbf{\bibinfo{volume}{17}}, \bibinfo{pages}{26--49} (\bibinfo{year}{2013}).

\bibitem{bib65}
\bibinfo{author}{Kravitz, D.~J.}, \bibinfo{author}{Saleem, K.~S.}, \bibinfo{author}{Baker, C.~I.} \& \bibinfo{author}{Mishkin, M.}
\newblock \bibinfo{title}{A new neural framework for visuospatial processing}.
\newblock \emph{\bibinfo{journal}{Nature Reviews Neuroscience}} \textbf{\bibinfo{volume}{12}}, \bibinfo{pages}{217--30} (\bibinfo{year}{2011}).

\bibitem{bib72}
\bibinfo{author}{Somerville, L.~H.} \emph{et~al.}
\newblock \bibinfo{title}{The lifespan human connectome project in development: A large-scale study of brain connectivity development in 5--21 year olds}.
\newblock \emph{\bibinfo{journal}{NeuroImage}} \textbf{\bibinfo{volume}{183}}, \bibinfo{pages}{456--468} (\bibinfo{year}{2018}).

\bibitem{bib43}
\bibinfo{author}{Smith, S.~M.}, \bibinfo{author}{Beckmann, C.~F.}, \bibinfo{author}{Andersson, J.}, \bibinfo{author}{Auerbach, E.~J.} \& \bibinfo{author}{Consortium, W. M.~H.}
\newblock \bibinfo{title}{Resting-state fmri in the human connectome project}.
\newblock \emph{\bibinfo{journal}{NeuroImage}} \textbf{\bibinfo{volume}{80}}, \bibinfo{pages}{144--168} (\bibinfo{year}{2013}).

\bibitem{bib71}
\bibinfo{author}{Bookheimer, S.~Y.} \emph{et~al.}
\newblock \bibinfo{title}{The lifespan human connectome project in aging: an overview}.
\newblock \emph{\bibinfo{journal}{NeuroImage}} \textbf{\bibinfo{volume}{185}}, \bibinfo{pages}{335--348} (\bibinfo{year}{2019}).

\bibitem{bib70}
\bibinfo{author}{Herculano-Houzel, S.} \emph{et~al.}
\newblock \bibinfo{title}{Connectivity-driven white matter scaling and folding in primate cerebral cortex}.
\newblock \emph{\bibinfo{journal}{Proceedings of the National Academy of Sciences}} \textbf{\bibinfo{volume}{107}}, \bibinfo{pages}{19008--19013} (\bibinfo{year}{2010}).

\bibitem{bib26}
\bibinfo{author}{Hamilton, W.~R.}
\newblock \emph{\bibinfo{title}{On a general method in dynamics}}  (\bibinfo{publisher}{Richard Taylor United Kindom}, \bibinfo{year}{1834}).

\bibitem{HUANG2024139}
\bibinfo{author}{Huang, S.}, \bibinfo{author}{De~Brigard, F.}, \bibinfo{author}{Cabeza, R.} \& \bibinfo{author}{Davis, S.~W.}
\newblock \bibinfo{title}{Connectivity analyses for task-based fmri}.
\newblock \emph{\bibinfo{journal}{Physics of life reviews}} \textbf{\bibinfo{volume}{49}}, \bibinfo{pages}{139--156} (\bibinfo{year}{2024}).

\bibitem{Jill2012Tools}
\bibinfo{author}{O’Reilly, J.~X.} \emph{et~al.}
\newblock \bibinfo{title}{Tools of the trade: psychophysiological interactions and functional connectivity}.
\newblock \emph{\bibinfo{journal}{Social Cognitive and Affective Neuroscience}} \textbf{\bibinfo{volume}{7}}, \bibinfo{pages}{604--609} (\bibinfo{year}{2012}).

\bibitem{2013Function}
\bibinfo{author}{Barch, D.~M.} \emph{et~al.}
\newblock \bibinfo{title}{Function in the human connectome: Task-fmri and individual differences in behavior}.
\newblock \emph{\bibinfo{journal}{NeuroImage}} \textbf{\bibinfo{volume}{80}}, \bibinfo{pages}{169--189} (\bibinfo{year}{2013}).

\bibitem{bib6}
\bibinfo{author}{Glasser, M.~F.} \emph{et~al.}
\newblock \bibinfo{title}{A multi-modal parcellation of human cerebral cortex}.
\newblock \emph{\bibinfo{journal}{Nature}} \textbf{\bibinfo{volume}{536}}, \bibinfo{pages}{171--178} (\bibinfo{year}{2016}).

\bibitem{bib33}
\bibinfo{author}{Daume, J.} \emph{et~al.}
\newblock \bibinfo{title}{Control of working memory by phase–amplitude coupling of human hippocampal neurons}.
\newblock \emph{\bibinfo{journal}{Nature}} \textbf{\bibinfo{volume}{629}}, \bibinfo{pages}{393--401} (\bibinfo{year}{2024}).

\bibitem{2016TheHamiltonianBrain}
\bibinfo{author}{Laurence, A.}, \bibinfo{author}{Máté, L.} \& \bibinfo{author}{Kording, K.~P.}
\newblock \bibinfo{title}{The hamiltonian brain: Efficient probabilistic inference with excitatory-inhibitory neural circuit dynamics}.
\newblock \emph{\bibinfo{journal}{PLoS Computational Biology}} \textbf{\bibinfo{volume}{12}} (\bibinfo{year}{2016}).

\bibitem{SENGUPTA20161107}
\bibinfo{author}{Sengupta, B.}, \bibinfo{author}{Friston, K.~J.} \& \bibinfo{author}{Penny, W.~D.}
\newblock \bibinfo{title}{Gradient-based mcmc samplers for dynamic causal modelling}.
\newblock \emph{\bibinfo{journal}{NeuroImage}} \textbf{\bibinfo{volume}{125}}, \bibinfo{pages}{1107--1118} (\bibinfo{year}{2016}).

\bibitem{baldy2025dynamic}
\bibinfo{author}{Baldy, N.}, \bibinfo{author}{Woodman, M.}, \bibinfo{author}{Jirsa, V.~K.} \& \bibinfo{author}{Hashemi, M.}
\newblock \bibinfo{title}{Dynamic causal modelling in probabilistic programming languages}.
\newblock \emph{\bibinfo{journal}{Journal of the Royal Society Interface}} \textbf{\bibinfo{volume}{22}}, \bibinfo{pages}{20240880} (\bibinfo{year}{2025}).

\bibitem{2015electromechanical}
\bibinfo{author}{Drapaca, C.~S.}
\newblock \bibinfo{title}{An electromechanical model of neuronal dynamics using hamilton's principle}.
\newblock \emph{\bibinfo{journal}{Frontiers in Cellular Neuroscience}} \textbf{\bibinfo{volume}{9}}, \bibinfo{pages}{271} (\bibinfo{year}{2015}).

\bibitem{Deco2025}
\bibinfo{author}{Deco, G.}, \bibinfo{author}{Perl, Y.~S.} \& \bibinfo{author}{Kringelbach, M.~L.}
\newblock \bibinfo{title}{Non-local schrödinger diffusion model reveals mechanisms of critical brain dynamics}.
\newblock \emph{\bibinfo{journal}{Cell Reports Physical Science}} \textbf{\bibinfo{volume}{6}}, \bibinfo{pages}{102663} (\bibinfo{year}{2025}).

\bibitem{bib30}
\bibinfo{author}{Abraham, A.} \emph{et~al.}
\newblock \bibinfo{title}{Machine learning for neuroimaging with scikit-learn}.
\newblock \emph{\bibinfo{journal}{Frontiers in Neuroinformatics}} \textbf{\bibinfo{volume}{8}}, \bibinfo{pages}{14} (\bibinfo{year}{2014}).

\bibitem{bib49}
\bibinfo{author}{Perl, Y.~S.} \emph{et~al.}
\newblock \bibinfo{title}{Data augmentation based on dynamical systems for the classification of brain states}.
\newblock \emph{\bibinfo{journal}{Chaos, Solitons \& Fractals}} \textbf{\bibinfo{volume}{139}}, \bibinfo{pages}{110069} (\bibinfo{year}{2020}).

\bibitem{effec_connec}
\bibinfo{author}{Rolls, E.~T.}, \bibinfo{author}{Gustavo, D.}, \bibinfo{author}{Chu-Chung, H.} \& \bibinfo{author}{Jianfeng, F.}
\newblock \bibinfo{title}{The effective connectivity of the human hippocampal memory system}.
\newblock \emph{\bibinfo{journal}{Cerebral Cortex}} \bibinfo{pages}{17} (\bibinfo{year}{2022}).

\bibitem{2022Asymmetric}
\bibinfo{author}{Parkes, L.},  \emph{et~al.}
\newblock \bibinfo{title}{Asymmetric signaling across the hierarchy of cytoarchitecture within the human connectome}.
\newblock \emph{\bibinfo{journal}{Science Advances}} \textbf{\bibinfo{volume}{8}}, \bibinfo{pages}{eadd2185} (\bibinfo{year}{2022}).

\bibitem{Nartall}
\bibinfo{author}{Nartallo-Kaluarachchi, R.} \emph{et~al.}
\newblock \bibinfo{title}{Multilevel irreversibility reveals higher-order organization of nonequilibrium interactions in human brain dynamics}.
\newblock \emph{\bibinfo{journal}{Proceedings of the National Academy of Sciences}} \textbf{\bibinfo{volume}{122}}, \bibinfo{pages}{e2408791122} (\bibinfo{year}{2025}).

\bibitem{2025Non}
\bibinfo{author}{Geli, S.~M.}, \bibinfo{author}{Lynn, C.~W.}, \bibinfo{author}{Kringelbach, M.~L.}, \bibinfo{author}{Deco, G.} \& \bibinfo{author}{Perl, Y.~S.}
\newblock \bibinfo{title}{Non-equilibrium whole-brain dynamics arise from pairwise interactions}.
\newblock \emph{\bibinfo{journal}{Cell Reports Physical Science}} \textbf{\bibinfo{volume}{6}} (\bibinfo{year}{2025}).

\bibitem{sun2025lifespan}
\bibinfo{author}{Sun, L.}, \bibinfo{author}{Zhao, T.}, \bibinfo{author}{Liang, X.} \emph{et~al.}
\newblock \bibinfo{title}{Human lifespan changes in the brain’s functional connectome}.
\newblock \emph{\bibinfo{journal}{Nature Neuroscience}} \textbf{\bibinfo{volume}{28}}, \bibinfo{pages}{891--901} (\bibinfo{year}{2025}).

\bibitem{Subtypes}
\bibinfo{author}{Danan}, \bibinfo{author}{Gu} \emph{et~al.}
\newblock \bibinfo{title}{Concordance and discordance of self-rated and researcher-measured successful aging: Subtypes and associated factors}.
\newblock \emph{\bibinfo{journal}{Journals of Gerontology}} \textbf{\bibinfo{volume}{72}}, \bibinfo{pages}{214--227} (\bibinfo{year}{2017}).

\bibitem{bib73}
\bibinfo{author}{Farahani, A.} \emph{et~al.}
\newblock \bibinfo{title}{Cerebral blood perfusion across biological systems and the human lifespan}.
\newblock \emph{\bibinfo{journal}{bioRxiv}}  (\bibinfo{year}{2025}).
\newblock \urlprefix\url{https://doi.org/10.1101/2025.02.05.636674}.

\bibitem{2023Publisher}
\bibinfo{author}{Matej, P.} \emph{et~al.}
\newblock \bibinfo{title}{Publisher correction: Functional brain networks in the evaluation of patients with neurodegenerative disorders}.
\newblock \emph{\bibinfo{journal}{Nature reviews. Neurology}} \textbf{\bibinfo{volume}{19}}, \bibinfo{pages}{384--384} (\bibinfo{year}{2023}).

\bibitem{0The}
\bibinfo{author}{Fornito, A.}, \bibinfo{author}{Zalesky, A.} \& \bibinfo{author}{Breakspear, M.}
\newblock \bibinfo{title}{The connectomics of brain disorders}.
\newblock \emph{\bibinfo{journal}{Nature Reviews Neuroscience}} \textbf{\bibinfo{volume}{16}}, \bibinfo{pages}{159--172} (\bibinfo{year}{2015}).

\bibitem{huang2022extended}
\bibinfo{author}{Huang, C.} \& \bibinfo{author}{et~al.}
\newblock \bibinfo{title}{An extended human connectome project multimodal parcellation atlas of the human cortex and subcortical areas}.
\newblock \emph{\bibinfo{journal}{NeuroImage}} \textbf{\bibinfo{volume}{244}}, \bibinfo{pages}{118589} (\bibinfo{year}{2022}).

\bibitem{2016The}
\bibinfo{author}{Lingzhong, F.} \emph{et~al.}
\newblock \bibinfo{title}{The human brainnetome atlas: A new brain atlas based on connectional architecture}.
\newblock \emph{\bibinfo{journal}{Cerebral Cortex}} \textbf{\bibinfo{volume}{26}}, \bibinfo{pages}{3508--3526} (\bibinfo{year}{2016}).

\bibitem{2013Groupwise}
\bibinfo{author}{Shen, X.}, \bibinfo{author}{Tokoglu, F.}, \bibinfo{author}{Papademetris, X.} \& \bibinfo{author}{Constable, R.~T.}
\newblock \bibinfo{title}{Groupwise whole-brain parcellation from resting-state fmri data for network node identification}.
\newblock \emph{\bibinfo{journal}{NeuroImage}} \textbf{\bibinfo{volume}{82}}, \bibinfo{pages}{403--415} (\bibinfo{year}{2013}).

\bibitem{bib41}
\bibinfo{author}{Horel, J.~D.}
\newblock \bibinfo{title}{Complex principal component analysis: Theory and examples}.
\newblock \emph{\bibinfo{journal}{Journal of Climatology \& Applied Meteorology}} \textbf{\bibinfo{volume}{23}}, \bibinfo{pages}{1660--1673} (\bibinfo{year}{1984}).

\bibitem{2019Cycle}
\bibinfo{author}{Cole, S.} \& \bibinfo{author}{Voytek, B.}
\newblock \bibinfo{title}{Cycle-by-cycle analysis of neural oscillations}.
\newblock \emph{\bibinfo{journal}{Journal of Neurophysiology}} \textbf{\bibinfo{volume}{122}}, \bibinfo{pages}{849--861} (\bibinfo{year}{2019}).

\bibitem{0Time}
\bibinfo{author}{Munia, T. T.~K.} \& \bibinfo{author}{Aviyente, S.}
\newblock \bibinfo{title}{Time-frequency based phase-amplitude coupling measure for neuronal oscillations}.
\newblock \emph{\bibinfo{journal}{Scientific Reports}} \textbf{\bibinfo{volume}{9}}, \bibinfo{pages}{12441} (\bibinfo{year}{2019}).

\bibitem{bib74}
\bibinfo{author}{Manea, A.~M.} \emph{et~al.}
\newblock \bibinfo{title}{Intrinsic timescales as an organizational principle of neural processing across the whole rhesus macaque brain}.
\newblock \emph{\bibinfo{journal}{eLife}} \textbf{\bibinfo{volume}{11}}, \bibinfo{pages}{e75540} (\bibinfo{year}{2020}).

\bibitem{bib75}
\bibinfo{author}{Ito, T.}, \bibinfo{author}{Hearne, L.~J.} \& \bibinfo{author}{Cole, M.~W.}
\newblock \bibinfo{title}{A cortical hierarchy of localized and distributed processes revealed via dissociation of task activations, connectivity changes, and intrinsic timescales}.
\newblock \emph{\bibinfo{journal}{NeuroImage}} \textbf{\bibinfo{volume}{221}}, \bibinfo{pages}{117141} (\bibinfo{year}{2020}).

\bibitem{bib36}
\bibinfo{author}{Wang, R.} \emph{et~al.}
\newblock \bibinfo{title}{Hierarchical connectome modes and critical state jointly maximize human brain functional diversity}.
\newblock \emph{\bibinfo{journal}{Physical Review Letters}} \textbf{\bibinfo{volume}{123}}, \bibinfo{pages}{038301} (\bibinfo{year}{2019}).

\end{thebibliography}
% common bib file
%% if required, the content of .bbl file can be included here once bbl is generated
%%\input sn-article.bbl

\newpage

\makeatletter
\renewcommand{\fnum@figure}{Extended Data Fig.\arabic{figure}}
\makeatother
\setcounter{figure}{0} 

\begin{figure*}[htbp]
	\centering
	\includegraphics[width=1\textwidth]{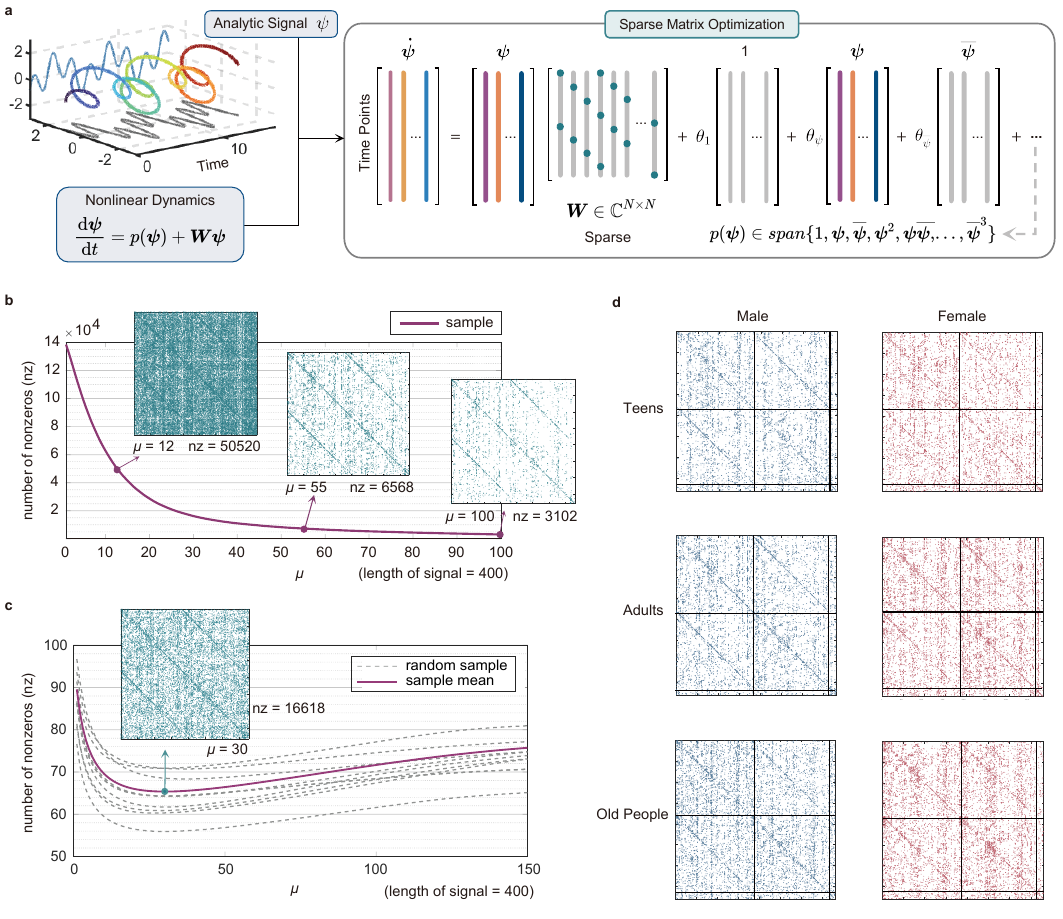} %
 	\caption{$\bm{\mathrm{a}}$, A complex-valued nonlinear model utilizes a generic form, with its parameters estimated through sparse matrix optimization. $\bm{\mathrm{b}}$, $\mu$ governs the sparsity of the resting-state coupling matrix.  $\bm{\mathrm{c}}$, Cross-validated parameter optimization is performed. The sparsity parameter ($\mu = 30$) is determined using training data from 30 participants (400 TRs per participant), demonstrating high inter-subject consistency. $\bm{\mathrm{d}}$, The pattern of nonzero elements in the coupling matrix is consistent across age and gender.  } 
	\label{fig:7}
\end{figure*}

\begin{figure*}[htpb]
	\centering
	\includegraphics[width=1\textwidth]{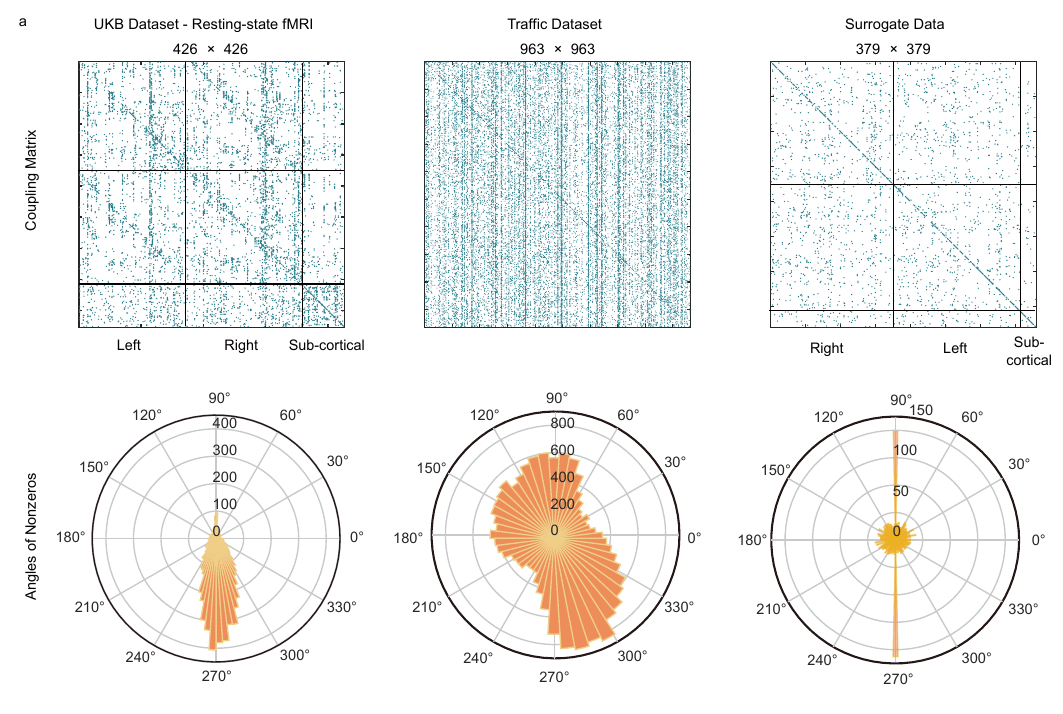} %
 	\caption{We derive the governing equations from the UK Biobank resting-state fMRI data (426 regions), New York City (NYC) traffic flow networks, and surrogate data from the Human Connectome Project (HCP). The topology of the coupling matrix (top) and the distribution of the angles of its non-zero elements (bottom) reveal distinct dynamical regimes: (i) UK Biobank dataset: Schr\"{o}dinger dynamics characterized by a tridiagonal coupling structure and a bimodal distribution of angles (1365 subjects; peaks at $-\pi/2$ and $\pi/2$, $R=0.519$, $p=0$, Rayleigh test after angle doubling). (ii) NYC traffic: single-diagonal structure with nonzero elements deviating from purely imaginary values, exhibiting weak phase locking ($R=0.045$). (iii), Surrogate HCP data: single-diagonal structure with minimal phase synchronization ($R=0.004$). } 
	\label{fig:surrogate}
\end{figure*}

\begin{figure*}[htbp]
	\centering
	\includegraphics[width=1\textwidth]{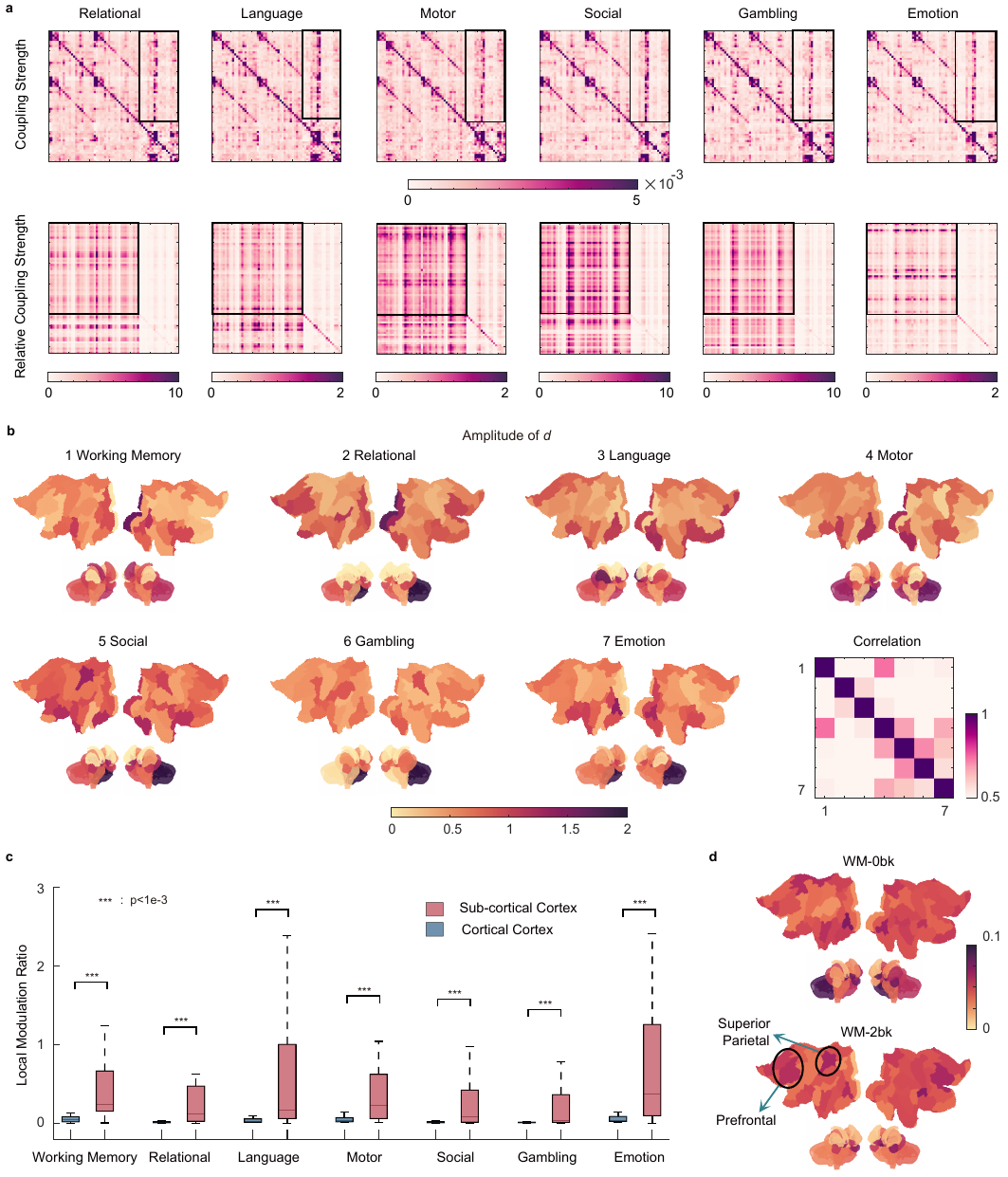} %
 	\caption{Details of task-induced coupling modulation. $\bm{\mathrm{a}}$, The coupling strength and relative coupling strength between brain regions are analyzed across different tasks. $\bm{\mathrm{b}}$, The diagonal factor $\bm{d}$ represents the local scaling of coupling strength for each region during task states. $\bm{\mathrm{c}}$, The local modulation ratio $(d_i / v_i)$ is significantly higher in subcortical regions than in the cerebral cortex across various tasks, indicating that the local rescaling $\bm{d}$ predominantly influences coupling in subcortical regions.  $\bm{\mathrm{d}}$, When comparing the coupling modulation factors between the 0-back and 2-back working memory tasks, we find that the 2-back condition is characterized by significantly stronger dynamic receptive tuning ($\bm{u}$) in the prefrontal and parietal cortices.} 
	\label{fig:task_appendix}
\end{figure*}

\end{document}